\documentclass[letterpaper,twocolumn,10pt]{article}
\usepackage{usenix}
\usepackage[available]{usenixbadges}

\usepackage{titlesec}
\titlespacing*{\section}{0pt}{7pt plus 2pt minus 1pt}{4pt plus 1pt}
\titlespacing*{\subsection}{0pt}{5pt plus 2pt minus 1pt}{3pt plus 1pt}

\usepackage{booktabs}  % for tables
\usepackage{dcolumn}   % for tables
\usepackage{multirow}  % for tables
\usepackage{colortbl}  % colorful columns and rows
\usepackage{array}     % needed for 'b' argument in tabular preamble
\usepackage{rotating}
\usepackage{xspace}
\usepackage{hyphenat}
\usepackage{enumerate}
\usepackage{comment}
\usepackage{balance}
\usepackage{listings}
\usepackage{framed}
\usepackage{balance}
\usepackage{amssymb,amsfonts,amsthm,amsmath,amscd}

\PassOptionsToPackage{hyphens}{url}
\usepackage{xurl}

\usepackage{hyperref}
\hypersetup{
  colorlinks=true,
  linkcolor=blue!90!red,
}

\usepackage[most]{tcolorbox}
\usepackage{xcolor}

\definecolor{bluepurple}{HTML}{C2C2FF}
\newcommand{\lstbg}[3][0pt]{{\fboxsep#1\colorbox{#2}{\strut #3}}}
\lstdefinelanguage{diffcpp}{
  language=C++,
  basicstyle=\scriptsize\ttfamily,
  columns=fullflexible,
  aboveskip=0pt,
  belowskip=0pt,
  lineskip=0pt,
  morecomment=[f][\lstbg{red!20}]{-},
  morecomment=[f][\lstbg{green!20}]{+},
  morecomment=[f][\lstbg{bluepurple}]{@@},
}

\usepackage[inline,shortlabels]{enumitem}
\usepackage{tikz}

\usepackage{wrapfig}
\usepackage{textcomp}
\usepackage{tabularx}
\usepackage{pifont}

\usepackage{wrapfig}
\usepackage{mdframed}
\usepackage{caption}
\usepackage{subcaption}

\usepackage{ulem} % for strikethrough and underlining
\usepackage{dblfloatfix}

\usepackage{makecell} % for pseudocode
\usepackage{algorithm}
\usepackage{algorithmicx}

\newif\ifextended
\newif\iflongbatching
\newif\ifsubmission
\newif\ifelementary

\ifx\buildextended\undefined
\else
    \extendedtrue
\fi

\ifx\noeditingmarks\undefined
   \newcommand{\pgwrapper}[3]{\begingroup \color{#1} #2: #3 \endgroup}
   \newcommand{\pgwrapperb}[1]{\textbf{#1}}
\else
   \newcommand{\pgwrapperb}[1]{}
   \newcommand{\pgwrapper}[3]{}
\fi

\newcommand{\heading}[1]{
\vspace{1ex}
\noindent
\textbf{#1}
}

\def\hn{\usefont{OT1}{phv}{mc}{n}\selectfont}

\newcommand{\mpfont}{\hn\scriptsize}

\ifx\noeditingmarks\undefined
    \newcommand{\MPworker}[2]{{\color{#1}\vrule\vrule}{\marginpar{\color{#1}\mpfont #2}}}
\else
    \newcommand{\MPworker}[2]{}
\fi

\makeatletter
\renewcommand*{\@fnsymbol}[1]{\ensuremath{\ifcase#1\or \star\or \dagger\or \ddagger\or
   \mathsection\or \mathparagraph\or \|\or **\or \dagger\dagger
   \or \ddagger\ddagger \else\@ctrerr\fi}}
\makeatother

\def\compactify{\itemsep=0in \topsep=2pt \parsep=0.00in \partopsep=0pt
\leftmargin=2em}
\let\latexusecounter=\usecounter

\newenvironment{myitemize}%
  {\begin{list}{\labelitemi}{\itemsep3pt \topsep3pt \parsep0.00in
  \partopsep=3pt \leftmargin1.2em}}%
  {\end{list}}
\newenvironment{myitemize2}%
  {\begin{list}{\labelitemi}{\itemsep1pt \topsep2pt \parsep0.00in
  \partopsep=0pt \leftmargin1.2em}}%
  {\end{list}}
  {\begin{list}{\labelitemi}{\itemsep2pt \topsep2pt \parsep0.00in
  \partopsep=0pt \leftmargin1.2em}}%
  {\end{list}}
  {\begin{list}{\threequartdash}{\itemsep3pt \topsep3pt \parsep0.00in
  \partopsep=3pt \leftmargin1.5em}}%
  {\end{list}}

\newenvironment{myenumerate}
  {\def\usecounter{\compactify\latexusecounter}
   \begin{enumerate}}
  {\end{enumerate}\let\usecounter=\latexusecounter}

\def\compactsortof{\itemsep=0in \topsep=2pt \parsep=0.00in \partopsep=0pt
\leftmargin=1.7em}
\newenvironment{myenumerate2}
  {\def\usecounter{\compactsortof\latexusecounter}
   \begin{enumerate}}
  {\end{enumerate}\let\usecounter=\latexusecounter}

\def\compactsqueeze{\itemsep=0pt \topsep0pt \parsep=0ex \partopsep=0pt
\leftmargin=1.63em}

\ifx\normalpar\undefined
  
\else
  
\fi

\newcommand{\sys}{\textsc{Zeno}\xspace}
\newcommand{\expop}{expression operator\xspace}
\newcommand{\expops}{expression operators\xspace}
\newcommand{\id}{FID\xspace}
\newcommand{\ids}{FIDs\xspace}
\newcommand{\lut}{mapping store\xspace}

\newcommand{\cfmaps}{crypto-free mappings\xspace}
\def\ie{i.e.\xspace} 
\def\eg{e.g.\xspace}

\begin{document}

\title{\LARGE Accelerating Confidential Databases with Crypto-free Mappings}

\author{
{\rm Wenxuan Huang \qquad Zhanbo Wang \qquad Mingyu Li} \\[6pt]
Key Laboratory of System Software (Chinese Academy of Sciences) \\
Institute of Software, Chinese Academy of Sciences \\
University of Chinese Academy of Sciences
}

\maketitle
\pagestyle{empty}

\begin{abstract}

Confidential databases (CDBs) enable secure queries over sensitive data in untrusted cloud environments using confidential computing hardware. While adoption is growing, widespread deployment is hindered by high overheads from frequent synchronous cryptographic operations, which cause significant computational and I/O bottlenecks.

\sys is a novel CDB design that removes cryptographic operations from the critical path. It introduces \cfmaps that maintain data-independent identifiers within the database while securely mapping them to plaintext secrets in a trusted domain.
This paradigm shift yields substantial performance gains across industry-standard benchmarks (TPC-C, TPC-H) and a real-world industrial workload.
Specifically, \sys speeds up TPC-H queries by up to 53.1$\times$ on ARM S-EL2 and 94.7$\times$ on x86 TDX compared to HEDB.
\sys's optimization techniques have been integrated into GaussDB.

\end{abstract}

\section{Introduction}
\label{s:intro}

Cloud databases have become the backbone of modern digital infrastructure, with organizations increasingly outsourcing their data management to Database-as-a-Service (DBaaS) providers~\cite{amazon-aurora, azure-sql, google-sql}. However, this convenience comes with a critical security risk: cloud administrators have unrestricted access to all tenant data, leaving opportunities for both malicious and accidental exposure. Recent breaches, such as the Allianz Life exposure of 1.1 million personal records~\cite{data-breach-2025}, underscore this threat. For sensitive databases regulated by laws such as HIPAA~\cite{hippa} and PCI-DSS~\cite{pci-dss},
% (\eg, HIPAA-governed~\cite{hippa} medical databases and PCI-DSS-regulated~\cite{pci-dss} financial systems),
entrusting cloud infrastructure with plaintext data is untenable.

Confidential databases (CDBs) have emerged as a compelling solution. By leveraging trusted execution environments (TEEs), CDBs enable secure data upload and query execution while keeping data encrypted from the cloud provider.
Over the past two decades, researchers have explored various CDB designs~\cite{Bajaj2011TrustedDB, Arasu2013Cipherbase, Priebe2018EnclaveDB, Vinayagamurthy2019StealthDB, Fuhry2020EncDBDB}, and major cloud providers have deployed CDB systems~\cite{Antonopoulos2020AEv2, Wang2022Operon, Guo2021GaussDB, Li2023HEDB, yang2024secudb}.
% : Microsoft's Always Encrypted~\cite{Antonopoulos2020AEv2} protects Azure SQL, Alibaba's Operon~\cite{Wang2022Operon} secures daily financial workloads, and ByteDance's SecuDB~\cite{yang2024secudb} safeguards user data globally.
% These deployments signal growing industry recognition that confidential databases are transitioning from research prototype to production necessity.

\begin{figure}
\centering
\includegraphics[width=0.45\textwidth]{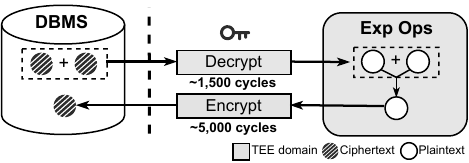}
\caption{In modern CDBs, a cross-domain invocation involves two decryptions and one encryption, incurring high computational overhead (CPU cycles evaluated on ARM).}
\label{fig:new-design}
\vspace{-1.0em}
\end{figure}

Despite this industry momentum, CDB systems face a critical barrier to widespread adoption:
severe performance degradation compared to plaintext databases, with analytical queries experiencing up to 79.5$\times$ higher latency~\cite{Li2023HEDB}.
This gap arises from a deliberate system design choice in industry deployments: rather than enclosing the entire DBMS within a TEE, which would prevent database administrators from performing essential maintenance tasks, modern CDBs execute only security-critical operators (\eg, numeric \texttt{SUM} operations) inside TEEs.
The resulting split architecture necessitates synchronous software-based decryption and encryption for every remote procedure call (RPC) across domains, as illustrated in \autoref{fig:new-design}. The computational burden scales linearly with data volume. For instance, a \texttt{SUM} aggregation over an $N$-row table requires 2$N$-2 decryptions and $N$-1 encryptions in total.
Even worse, to support arbitrary computation across fields, modern CDBs employ field-level encryption, inflating each 4-byte integer to 32 bytes (8$\times$ expansion) due to cryptographic metadata (a 12-byte nonce and a 16-byte authentication tag required by schemes like AES-GCM). This dramatically amplifies cross-domain data movement, memory pressure, and I/O bandwidth demands.
\iffalse
This architectural split creates a fundamental bottleneck: \emph{frequent, synchronous cryptographic operations on every cross-domain interaction}. For a simple \texttt{SUM} aggregation over an $N$-row table, each addition requires decrypting two operands, computing the sum, and re-encrypting the result before returning to the DBMS, as depicted in \autoref{fig:new-design}. For even modest tables with millions of rows---a common scale reported by DBaaS vendors we collaborate with---this translates to millions of encryption and decryption operations on the critical path.
Even worse, modern CDBs employ field-level encryption, where each encrypted field carries 28 bytes of metadata (12-byte nonce and 16-byte authentication tag required by schemes like AES-GCM), bloating an 4-byte integer to 32 bytes---an 8$\times$ expansion. This explosion amplifies every subsequent cost: more data copied across domains, more memory pressure, more I/O between buffer pools and storage.
\fi
These limitations raise a fundamental question: \textbf{Is cryptographic overhead inherent to CDB design, or can we achieve equivalent security guarantees with far cheaper alternatives?}

% For decades, the CDB community has assumed that any data leaving TEEs must be encrypted~\cite{Bajaj2011TrustedDB, Arasu2013Cipherbase, Priebe2018EnclaveDB, Vinayagamurthy2019StealthDB, Fuhry2020EncDBDB, Antonopoulos2020AEv2, Guo2021GaussDB, Wang2022Operon, Li2023HEDB, yang2024secudb}, necessitating expensive cryptographic mappings.
% Our work challenges this assumption.
% The insight is that when pointers (not data) are stored outside TEEs, TEE isolation alone provides sufficient confidentiality---the pointers reveal nothing about the referenced data; in essence, indirection can replace encryption.
\heading{Key insight.}
We argue that the performance crisis stems from an unexamined assumption that has persisted for decades: \emph{any data leaving TEEs must be encrypted.}
Modern CDBs store ciphertexts in the DBMS and decrypt them in TEEs for computation. We observe that this is fundamentally a mapping-based protection scheme: the DBMS uses ciphertexts as pointers to reference plaintext data in TEEs. % and that pointers need no encryption.
Current practice conflates indirection with protection, encrypting both pointers and data. We decouple these concerns: pointers use lightweight indirection; only data at rest uses encryption.
Based on this insight, we aim to replace the costly ciphertext-to-plaintext mappings with highly efficient index-to-plaintext mappings, while preserving both data confidentiality and database correctness.
% ---a 30-100$\times$ speedup per operation.
% Based on this insight, we replace the costly ciphertext-to-plaintext mappings with highly efficient \emph{\cfmaps}---lightweight, data-independent field identifiers (\ids) that still point to plaintext data stored in TEEs. This preserves both database correctness and data confidentiality while dramatically reducing overhead. Using simple integer \ids, the DBMS can reference data via low-latency put/get operations ($\sim$50 cycles) instead of encryption ($\sim$5,000 cycles) and decryption ($\sim$1,500 cycles)---a 30-100$\times$ speedup per operation.
% While intuitive in hindsight, the idea that cryptographic mappings are unnecessary contradicts conventional wisdom in CDB design.

\begin{table}[t]
\centering
\caption{Comparison between modern CDBs and \sys (CPU cycles evaluated on ARM).}
\label{fig:comparison}
\vspace{-0.6em}
\scalebox{0.84}{
\begin{tabular}{@{}lll@{}}
\toprule
& \textbf{Modern CDBs} & \textbf{\sys} \\
\midrule
\textbf{Indirection} & Ciphertexts & Data-independent \ids \\
\textbf{Protection} & Ciphertexts & Encryption at rest \\
% \textbf{Indirection cost} & \makecell[l]{Decrypt ($\sim$1,500 cycles) + \\ Encrypt ($\sim$5,000 cycles)} & Get/Put ($\sim$80 cycles) \\
\textbf{Indirection cost} 
& \makecell[l]{Decrypt ($\sim$1,500 cycles) \\ Encrypt ($\sim$5,000 cycles)} 
& \makecell[l]{Get ($\sim$30 cycles) \\ Put ($\sim$200 cycles)} \\
% \textbf{Coupling} & Conflated & Decoupled \\
\bottomrule
\end{tabular}
}
\end{table}

\heading{Our approach.}
We present \sys, a CDB system that eliminates compute-intensive cryptography from the critical path without sacrificing security.
\sys introduces \emph{\cfmaps} that bind data-independent field identifiers (\ids) to their corresponding plaintext fields within TEEs. This allows the DBMS to manipulate data as before, while
replacing slow per-field en/decryption with ultra-low-latency put/get operations, as shown in \autoref{fig:comparison}.
% In addition, \sys eliminates synchronous per-field encryption. When evicting data from TEE memory to untrusted storage, \sys applies asynchronous block-level encryption instead of fine-grained field-level encryption.
In addition, to avoid encrypting every field synchronously, \sys defers encryption: when data must be evicted from TEE memory to untrusted storage, \sys performs asynchronous block-level encryption rather than fine-grained field-level encryption.
In short, \cfmaps deliver two key benefits: they remove cryptographic operations from query execution hot paths and reduce metadata overhead via coarser-grained encryption.

Our design brings two main challenges.
First,
% increasing dataset size could cause the mapping lookup time to grow, which may ultimately become more costly than the cryptographic operations themselves.
as datasets scale, the mapping lookup time could grow unbounded and ultimately exceed the cost of constant-time cryptographic operations.
Second, maintaining transaction semantics and mapping consistency requires new mechanisms because both domains are now stateful, a situation not encountered in conventional CDB systems.

To overcome the first challenge, \sys introduces a scalable, low-latency data structure called the \emph{\lut}. This \lut provides $O(1)$ access complexity by treating \ids as direct array indices, eliminating the need for tree traversals or hash-table probing. \sys further preserves data locality across domains through storage-layout-aware \lut organization. This layout awareness, combined with proactive prefetching, keeps the corresponding data resident in TEE memory when the DBMS accesses an \id block, thereby minimizing additional I/O overhead.

The second challenge requires a subtle observation: the state of the \lut is internal to the system and invisible to the DBMS. The DBMS observes only the \ids it maintains---the system's external state. This asymmetry allows us to adopt external synchrony~\cite{nightingale2008rethink}: only \ids maintained by the DBMS should be mapped to valid, up-to-date plaintexts; the reverse mapping (plaintexts to \ids) need not even exist. As a result, \sys only needs to maintain the invariant that every \id in the DBMS has a corresponding valid value in the \lut. To uphold this invariant across transactions and crashes, \sys introduces lightweight coordination protocols that ensure cross-domain updates remain consistent and crash-tolerant with minimal overhead.

We evaluate \sys against HEDB~\cite{Li2023HEDB}, a state-of-the-art modern CDB that achieves the trifecta of functionality, maintainability, and security.
Our results across both ARM S-EL2 and x86 TDX show that \sys achieves substantial performance improvements while preserving all desired properties.
On the transactional benchmark TPC-C, \sys reduces HEDB's throughput loss relative to plaintext PostgreSQL by up to 49.8\% on ARM S-EL2 and 73.8\% on x86 TDX. On the analytical benchmark TPC-H, \sys speeds up query execution by up to 53.1$\times$ on ARM S-EL2 and 94.7$\times$ on x86 TDX over HEDB. On real-world industrial queries, \sys outperforms HEDB with average speedups of 4.6$\times$ on ARM S-EL2 and 2.9$\times$ on x86 TDX.
Additionally, \sys reduces storage consumption by 38.9\%, 52.8\%, and 42.4\% across the three workloads, respectively.
%%%%% Compared to the plaintext vanilla database, \sys achieves 45.9-91.3\% throughput in transactional workloads and incurs 0.3-23.6$\times$ overhead in analytical workloads. This indicates a performance gap in certain cases, reflecting the cost of security.
The main performance techniques of \sys have been integrated into GaussDB.

\heading{Contributions.}
We make the following contributions:
\begin{myitemize2}
\item We identify and analyze the performance bottlenecks of modern CDBs through comprehensive profiling and empirical analysis.
\item We propose a novel CDB design with \cfmaps, which remove costly cryptographic operations from the critical path of existing modern CDBs.
\item We build and deploy \sys, which outperforms state-of-the-art modern CDBs across transactional and analytical benchmarks, as well as a real-world industrial workload.
\end{myitemize2}

\section{Background and Motivation}
\label{s:motivation}
\label{s:background}

\subsection{Confidential Computing}

Confidential computing~\cite{amd-sev, intel-sgx, intel-tdx, arm-cca, aws-nitro, lee2020keystone, feng2021penglai, Li2021TwinVisor, Hunt2021PEF, Li2022CCA} provides hardware-based trusted execution environments (TEEs) with three security guarantees. First, secure isolation through either enclaves~\cite{intel-sgx, feng2021penglai, lee2020keystone} or confidential virtual machines (CVMs)~\cite{amd-sev, intel-tdx, arm-cca} protects against inspection or tampering by untrusted privileged software. Second, hardware root-of-trust enables clients to verify code authenticity via remote attestation. Third, memory encryption and optional integrity checks resist physical attacks. Confidential computing is now widely available on commercial clouds~\cite{google-sev, azure-sgx, aws-nitro, ibm-hyperprotect, alibaba-cc}.

\subsection{Confidential Databases}
\label{s:cdb-bg}

Database-as-a-Service (DBaaS)~\cite{Hacigumus2002DBaaS} offers fully managed databases that handle updates, backups, and maintenance for users. However, since cloud environments are untrusted, sensitive data like healthcare and financial information requires secure storage and processing.
Confidential databases (CDBs)~\cite{Bajaj2011TrustedDB, Arasu2013Cipherbase, Priebe2018EnclaveDB, Ribeiro2018DBStore, Vinayagamurthy2019StealthDB, Fuhry2020EncDBDB, Li2023HEDB, Antonopoulos2020AEv2, Guo2021GaussDB, Wang2022Operon, yang2024secudb} leverage confidential computing to enable SQL query processing over encrypted data without compromising confidentiality.

Existing CDB systems can be broadly categorized into two types.
One type~\cite{Bajaj2011TrustedDB, Priebe2018EnclaveDB, Ribeiro2018DBStore, yang2024secudb} places the entire DBMS inside the trusted domain (TEE), where the DBMS processes data in plaintext and encrypts it when interacting with untrusted environments (\eg, network or storage).
However, this type imposes an inherent limitation: the black-box TEE hinders DBMS maintainability for database administrators (DBAs)~\cite{Antonopoulos2020AEv2, Li2023HEDB}. % since granting them access to the TEE would expose sensitive data
In contrast, another type~\cite{Vinayagamurthy2019StealthDB, Antonopoulos2020AEv2, Guo2021GaussDB, Wang2022Operon, Li2023HEDB} adopts a split architecture: only essential expression operators (\eg, arithmetic, comparison\footnote{All operators yield encrypted results, except for comparisons (\eg, $<$, $=$, $>$), which output plaintext boolean values for use in indexing ciphertexts.}) execute inside TEEs, while most DBMS logic (\eg, SQL parsing, planning, buffer management) remains in the untrusted domain. This minimizes the TCB and preserves maintainability: DBAs can inspect query plans and debug DBMS issues. We refer to the latter as \emph{``modern CDBs''} throughout the paper.

To enable flexible computation across arbitrary fields, modern CDBs apply field-level encryption.
% After remote attestation, users submit queries that the DBMS executes by invoking TEE operators.
For example, since all fields from the column \texttt{Salary} are encrypted, \texttt{"SELECT SUM(Salary) FROM Employee"} requires the DBMS to iteratively call the \texttt{addition} operator inside the TEE, which decrypts operands, computes the sum, re-encrypts the result, and returns it to the DBMS.
While secure, this design incurs substantial performance overhead, as we will analyze later.

\heading{HEDB~\cite{Li2023HEDB}: a representative modern CDB.}
Modern CDBs could be vulnerable to \emph{smuggle attacks}~\cite{Li2023HEDB}: malicious DBAs exploit operator interfaces to construct known ciphertexts (via arithmetic operators, \eg, division yields encrypted ``one,'' addition builds sequences) and compare them against victim ciphertexts (via equality operators), thereby extracting secrets without authorization.
HEDB addresses smuggle attacks through a dual-TEE architecture: an \emph{integrity zone} that hosts the DBMS and a \emph{privacy zone} that hosts operators, with a secure channel preventing direct interface access by DBAs.
For maintainability, HEDB records cross-zone calls as ciphertext, enabling DBAs to replay and debug queries without plaintext access.
% \TODO{REVIEW: generality to other Type-II CDB}
Therefore, \sys builds on HEDB's security foundation.
Notably, the performance limitations addressed by \sys exist across many split-architecture CDBs~\cite{Vinayagamurthy2019StealthDB, Antonopoulos2020AEv2, Guo2021GaussDB, Wang2022Operon}, and its \cfmaps generalize beyond HEDB.
% Modern CDBs remain vulnerable to interface attacks by malicious insiders.
% To counter interface attacks while preserving maintainability, HEDB employs a dual-TEE architecture: an \emph{integrity zone} that hosts the DBMS for SQL execution, and a \emph{privacy zone} that hosts expression operators for computation over secrets. Two zones communicate only via a secure channel, preventing privileged attackers (\eg, DBAs) from exploiting operator interfaces to extract sensitive data.
% For maintainability, all cross-zone calls are recorded using ciphertext, enabling DBAs to replay and debug queries offline without accessing plaintext.
% \sys inherits and improves this design, but but remains compatible with any modern CDB, not only HEDB.
% HEDB records all cross-zone invocations in encrypted form, enabling DBAs to replay and debug queries in an untrusted environment without accessing plaintext data.
% This dual-zone architecture provides the security and maintainability properties that \sys inherits and optimizes. Note that \sys is not limited to HEDB, but for all modern CDBs.

\subsection{What Makes Modern CDB Systems Slow?}
\label{s:motivation_analysis}

Modern CDBs exemplified by HEDB have demonstrated significant performance overheads. Benchmarks on TPC-H show that HEDB can incur up to a 79.5$\times$ increase in latency compared to a plaintext database~\cite{Li2023HEDB}.
These overheads render modern CDBs impractical for latency-sensitive applications, hindering broader industry adoption.
% Our analysis pinpoints two factors.
Our analysis attributes them to two main factors.

\heading{Factor-\#1: expensive encryption and decryption.}
In modern CDBs, each computation on sensitive data requires a cross-domain round-trip, which involves two decryptions and one encryption, each taking thousands of cycles, as shown in~\autoref{fig:comparison}.
For an operation like \texttt{SUM} on a table with $N$ rows, the computation incurs $N-1$ round-trips, resulting in $2N-2$ decryptions and $N-1$ encryptions. 
Given that $N$ is often \emph{tens of millions} in real-world scenarios (reported by several CDB vendors we surveyed), this high frequency of expensive cryptographic operations inevitably introduces high overhead on the critical path.
While batching multiple RPC invocations into a single round trip (\eg, for aggregation operators such as \texttt{SUM}, \texttt{AVG}, \texttt{MAX}, and \texttt{MIN}) can help amortize overhead, the core issue remains that cryptographic operations must still be performed on every single data item.

\heading{Factor-\#2: ciphertext expansion.}
Modern CDBs employ field-level encryption with AES-GCM.
To guarantee data confidentiality and integrity, each encrypted field requires a nonce (12 bytes) and message authentication code (MAC, 16 bytes), totaling 28 B of cryptographic metadata. This introduces significant ciphertext expansion: a 4-byte integer expands to 32 bytes, an 8$\times$ size increase.
This expansion creates two performance bottlenecks. First, it enlarges the data transferred during frequent inter-domain RPCs, resulting in higher memory copy latency. Second, it increases the overall storage footprint of the database, leading to more I/O operations between the DBMS buffer pool and storage.

\begin{figure}[t]
\centering
\includegraphics[width=0.48\textwidth]{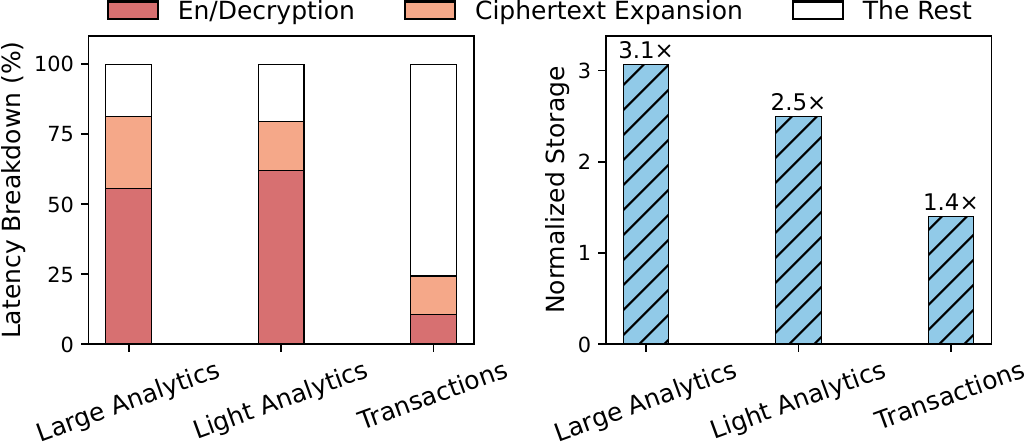}
\caption{Latency and storage overhead analysis across three typical workloads.
\textbf{Left (a)}: profiling end-to-end SQL execution time; the rest denotes normal DBMS execution and RPC invocations.
\textbf{Right (b)}: storage overhead normalized to plaintext databases.}
\label{fig:motivation-eval}
\end{figure}

\heading{Empirical analysis.}
We profile HEDB~\cite{Li2023HEDB} (commit f48742de) using three workloads:
(i) TPC-H Q1 with a dataset larger than memory, representing large table computation;
(ii) TPC-H Q1 with a dataset fitting in memory, capturing lightweight analytical settings; and 
(iii) a transactional benchmark with many write-heavy small queries.
Despite the batching optimization, HEDB exhibits significantly higher overhead, with an overall latency slowdown of up to 5.3$\times$ relative to the plaintext baseline.
% \autoref{fig:motivation-eval} presents the breakdown of execution latency and storage costs.
As illustrated by the performance breakdown in \autoref{fig:motivation-eval}(a), en/decryption operations and ciphertext expansion account for a substantial fraction of the total execution time, contributing 10.5--62.6\% and 13.9--25.5\%, respectively, across workloads.
Specifically, the contribution of I/O amplification is more pronounced in (i) than in (ii). This shift shows that the overhead distribution varies with the workload, particularly with respect to data scale and memory residency.
Meanwhile, the storage overhead (\autoref{fig:motivation-eval}(b)) reveals that field-level encryption incurs a 1.4$\times$ to 3.1$\times$ storage expansion relative to the plaintext baseline.
% The remaining 8.6-26.5\% of the execution time is attributable to DBMS execution and RPC overhead across domains, the latter of which stems from frequent copying of expanded ciphertexts.

\heading{Strawman: a ciphertext-to-plaintext cache.}
Since confidential VM-based TEEs (\eg, Intel TDX~\cite{intel-tdx}, AMD SEV-SNP~\cite{amd-sev}) provide sufficiently large secure memory, one may consider maintaining a ciphertext-to-plaintext cache within TEEs to remove cryptographic operations from the critical path.
However, as the cache scales with the protected dataset, searching a large ciphertext-keyed structure can itself become a bottleneck.
As our ablation study shows (\S\ref{s:eval-ablation}), such a cache helps reduce cryptographic cost but remains limited by the overhead of ciphertext-based lookups.

\section{\sys Overview}
\label{s:overview}

\heading{Design goals.}
\sys has two design goals:
\begin{myitemize2}
    \item G1: \emph{Compatibility.}
    \sys aims to be compatible with modern CDB designs by preserving their functionality (standard SQL and ACID), while retaining HEDB-level DBA maintainability and security (data confidentiality and database integrity).
    \item G2: \emph{Efficiency.}
    \sys aims to reduce the performance gap between existing modern CDBs and plaintext databases by optimizing the two factors identified in \autoref{s:motivation_analysis}.
\end{myitemize2}

\heading{Threat model.}
% \TODO{PLEASE REVIEW HERE: Clarify leakage scope early: inherited functional leakage, FID-level representation metadata, and TEE-internal side channels}
We assume a powerful adversary who controls the entire cloud infrastructure except TEE hardware. This includes compromised server software, rogue cloud and database administrators, and external attackers. Adversaries can read and modify any data outside TEE domains---data at rest, in untrusted memory, and in transit---and observe all network traffic and I/O patterns visible outside TEE domains.

We assume TEEs provide two essential properties: \emph{memory isolation}, preventing external access to TEE-internal state, and \emph{remote attestation}, enabling users to verify code integrity before establishing trust. While TEE-agnostic, \sys requires these guarantees to ensure sensitive plaintext data remains visible only within TEE domains.
We do not consider TEE side-channel attacks (\eg, \cite{xu2015controlledchannel,van2024sok,schluter2024heckler,chuang2026teefail}); these can be addressed by orthogonal hardening techniques~\cite{zhou2023coreslicing, castes2024coregapping, Mavrogiannakis2025OBLIVIATOR}.
Nor do we consider database vulnerabilities~\cite{dbms-bugs}, TEE defects~\cite{Kocher2019Spectre}, physical attacks~\cite{Lee2020OffChip}, or denial-of-service.

Functional leakage inherent to split-architecture CDBs is permitted. This reflects a fundamental trade-off between functionality and privacy: exposing certain metadata is sometimes necessary to support core database functionalities, such as order for index construction~\cite{Li2023HEDB}.
Because our design substitutes ciphertexts with \ids, we additionally account for any metadata exposed through this indirection mechanism; its security implications are analyzed in \autoref{sec:security}.

\heading{Insights.}
% Current practice of modern CDBs uses a mappting-based protection scheme. They use ciphertexts as pointers. However, ciphertext-based pointers couple indirection and protection, which inevitably requires expensive decryption and encryption during cross-domain RPCs. \sys decouples indirection from protection, by using data-independent \ids for indirection, and encrypting data at rest for protection.
Current practice in modern CDBs employs a mapping-based protection scheme, using ciphertexts as pointers to plaintext data in TEEs. This design conflates two orthogonal concerns: \emph{indirection} (cross-domain references) and \emph{protection} (data confidentiality). Because ciphertexts serve both roles, this coupling necessitates expensive en/decryption on every RPC. \sys decouples them: \ids provide efficient indirection, while encryption protects data at rest.
% by using data-independent \ids for efficient indirection, and encrypting data at rest for protection.

In fact, separating indirection from protection is a long-standing systems principle.
Capability systems~\cite{dennis1980capability, levy2014capability} used unforgeable tokens protected by OS confinement, rather than cryptography. Similarly, OSes have long used opaque handles, \eg, Unix file descriptors~\cite{ritchie1974unix, accetta1986mach}, which are mapped internally by the kernel. In both cases, security comes from address-space isolation, keeping references simple and fast.

\begin{figure}[t]
\centering
\includegraphics[width=0.46\textwidth]{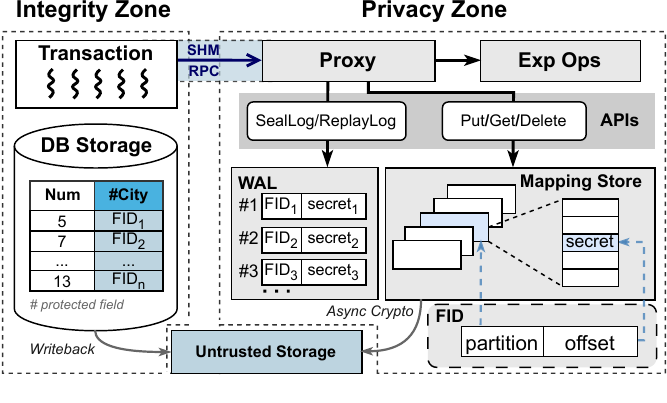}
\vspace{-0.5em}
\caption{High-level architecture of \sys.}
\label{fig:overview-arch}
\vspace{-1em}
\end{figure}

\heading{Architecture.}
\sys's high-level architecture is depicted in \autoref{fig:overview-arch}.
Following the security principles of HEDB's dual-zone design, \sys redesigns how protected values flow between zones.
% \sys inherits HEDB's dual-zone architecture but fundamentally redesigns how data flows between zones.
% While HEDB uses expensive cryptographic operations on every cross-zone interaction, \sys introduces \cfmaps that eliminate this bottleneck while preserving all security properties.
The integrity zone runs a commodity DBMS engine, and
the privacy zone comprises four components:
(1) a proxy that performs mapping operations and invokes the appropriate operators,
(2) \expops that perform computation over plaintexts,
(3) the \lut that manages the \id-to-secret mappings, and
(4) a write-ahead log (WAL) for cross-domain consistency.
The \lut is divided into temporary and permanent partitions for fine-grained management. % enhanced efficiency (\autoref{s:design:perf}).
% To ensure data durability and transaction consistency, \sys uses a commit protocol and a WAL for crash recovery. % (\autoref{s:design:txn}).
The proxy interacts with the remaining components through a set of well-defined APIs.

\heading{APIs.}\label{s:api}
\sys provides five APIs to manage the \lut: % (\autoref{tab:apis}).
% \sys maintains the \lut through the following APIs:
\begin{myitemize2}
    \item \texttt{Put(partition\_num, secret) $\rightarrow$ \id}:
    Stores the given secret in the specified partition of the \lut. Returns an \id with the partition number embedded, as a reference to the stored secret.
    \item \texttt{Get(\id) $\rightarrow$ secret}:
    Retrieves the secret associated with the specified \id, or returns a null value if the \id has no corresponding mapping.
    \item \texttt{Delete(\id)}:
    Invalidates the specified \id-to-secret mapping in the \lut, marking the entry as available for future allocation.
    \item \texttt{SealLog(records) $\rightarrow$ \texttt{encrypted\_records}}:
    Encrypts the given mapping-store WAL records and returns the encrypted records.
    \item \texttt{ReplayLog(\texttt{encrypted\_records})}:
    Replays the specified encrypted mapping-store WAL records to reconstruct \id-to-secret mappings.
\end{myitemize2}

At the operator interface, existing modern CDBs~\cite{Vinayagamurthy2019StealthDB, Antonopoulos2020AEv2, Guo2021GaussDB, Wang2022Operon, Li2023HEDB} can benefit from \sys's gains by replacing per-field en/decryption with \texttt{Put()}/\texttt{Get()} calls (see \autoref{lst:api-use}).

\begin{figure}[h]
\centering
\begin{lstlisting}[language=diffcpp, caption={Example code for extending the original expression operator using \sys's APIs.}, label={lst:api-use}, captionpos=b]
    /* expression operators (unchanged) */
    int int_add(int l, int r) { return l + r; }

    /* proxy code snippet */
    -    EncInt enc_int(Ops ops, EncInt left, EncInt right)
    +    FID enc_int(Ops ops, FID left, FID right)
    {
    -        int l = Decrypt(left), r = Decrypt(right);
    +        int l = Get(left), r = Get(right);
      int result;
      switch (ops) {
          case ADD: result = int_add(l, r); break;
          // other cases...
      }
    -        return Encrypt(result);
    +        // use temporary partitions for intermediate results
    +        return Put(TMP_PARTITION_NUM, result);
    }
\end{lstlisting}
\vspace{-1em}
\end{figure}

\heading{Workflow.}
\sys's query execution follows the workflow of modern CDB architectures. % (described in \autoref{s:cdb-bg}). 
To perform computation, the proxy fetches secrets associated with each input \id from the \lut via \texttt{Get()}, invokes the appropriate \expop to compute over these retrieved secrets, and then stores the result back to the \lut via \texttt{Put()}, which generates a new \id representing the output.
The DBMS proceeds with query execution using this returned \id.
% With the \lut in place, \sys can effectively bypass the en/decryption steps on the critical path.
Finally, the proxy resolves \ids back into ciphertexts by calling \texttt{Get()} followed by encrypting the retrieved results within the privacy zone before returning them to users.
Throughout this process, \ids remain opaque to users.
This design preserves \emph{compatibility} (G1) with the client-side encryption model used by modern CDBs~\cite{Antonopoulos2020AEv2, Guo2021GaussDB, Wang2022Operon, Li2023HEDB}.

\heading{Challenges.}
Since \sys introduces state into the privacy zone, it brings two challenges.
First, the lookup latency in the \lut may grow with dataset size and eventually exceed constant-time cryptographic operations (\autoref{s:design:perf}).
Second, maintaining state consistency between the integrity and privacy zones is necessary for database semantics (\autoref{s:design:txn}).

\section{Achieving Efficiency via Mapping Store}
\label{s:design:perf}

This section details the \lut, which plays a crucial role in achieving \emph{efficiency} (G2).
% \TODO{Should we add a subsection title: design rational?}

% To achieve \emph{efficiency} (G2), \sys addresses the two performance bottlenecks of modern CDBs (identified in \autoref{s:motivation_analysis}) by introducing the \lut to implement \cfmaps.
% Scaling the \cfmap to large-scale data volumes introduces non-trivial challenges.

% First, massive data volumes demand a scalable data structure to handle indexing (\texttt{Get()}) and modifications (\texttt{Put()} and \texttt{Delete()}) efficiently. Without such scalability, the overhead could eventually surpass that of cryptographic operations (\autoref{s:motivation_analysis}). % strawman approach
% We address this in \autoref{s:kv} through a high-performance structure.

% Second, \sys needs to be designed to be less sensitive to restricted memory resources than modern CDBs.
% To minimize the overall memory and storage footprint, we design mapping entries with much lower metadata overhead as detailed in \autoref{s:kv}.
% Furthermore, we overlap the I/O operations directed at the database and the \lut to prevent additional latency when \sys retrieves \ids and sensitive data from two separate domains, by enabling coordinated and efficient cross-domain data fetching, as detailed in \autoref{s:locality}.

\subsection{Mapping Management}
\label{s:kv}

\heading{Data structure design.}
The \lut is a linear array of tightly packed fields. Each sensitive field is assigned a unique integer \id that serves as a direct array index, hiding the size of individual fields from the adversary.
This design choice offers two advantages.
First, employing a compact integer \id effectively restricts the metadata overhead per field to a mere 8~B, representing a 71.4\% reduction compared to the 28~B metadata required by modern CDBs.
Second, the use of integer-based direct indexing enables $O(1)$, cache-friendly array access to the \lut, thereby eliminating the pointer-chasing and lookup overhead inherent in tree-based or hash-based key-value stores (\eg, \cite{leveldb, rocksdb, lmdb}).
% In contrast, systems that maintain global indices face higher contention due to frequent locking for consistency.

\heading{Mapping store operations.}
For fixed-length types (\eg, \texttt{FLOAT8}, \texttt{TIMESTAMP}), \texttt{Delete()} marks the corresponding mapping slots as reusable, and \texttt{Put()} either overwrites these invalidated slots or appends to the end.
For variable-length types (\eg, \texttt{VARCHAR}, \texttt{NUMERIC}), the situation is subtle because values cannot be overwritten in place.
We therefore maintain two arrays: an \emph{offset array} and a \emph{data array}. The offset array is indexed directly by \ids and stores pointers into the data array.
The data array is inspired by the slab allocator~\cite{bonwick1994slab}: fields are grouped into size-class buckets, each maintained as a contiguous packed array to minimize fragmentation under frequent allocation and deallocation. Within each bucket, insertions and deletions follow the same overwrite/append logic as the fixed-length case.
% and tracks them using a durable bitmap for future space reuse.
% To accelerate reuse, we maintain a free list as a cache of reusable \ids, which is periodically refreshed in the background.

\heading{Partition-based state management.}
The \lut cannot directly infer the lifetime of each field. For instance, when an \expop produces a result, it is unclear whether the value will be discarded after use or eventually persisted.
Careless reclamation risks leaving live \ids pointing to freed storage, causing data loss.
To address this, we classify values into two categories: ephemeral values (\eg, query constants and intermediate results) and persistent values (\eg, data that outlives query completion).
We separate their management by mapping them to distinct \lut partitions: a temporary partition for ephemeral values and permanent partitions for persistent values.

For ephemeral values, reclamation occurs throughout their lifecycle. They are first placed in the temporary partition via \texttt{Put()} with the temporary partition identifier \texttt{TMP\_PARTITION\_NUM}.
If the DBMS later decides to persist such a value, \sys intercepts the write path and performs another \texttt{Put()} with the permanent partition identifier.
During this \texttt{Put()} call, \sys allocates a new persistent \id, copies the value into the permanent partition, and returns the new \id to the DBMS.
To avoid accumulating stale values in the temporary partition, \sys reclaims ephemeral values in the privacy zone as soon as they are known to be dead, \eg, temporary operands consumed by aggregation or temporary \ids that have already been migrated to the permanent partition.
At the end of each query, any values still remaining in the temporary partition, \ie, ephemeral values whose lifetime extends to the query boundary, can be safely discarded.

For persistent values, reclamation is triggered when the DBMS instructs the proxy to invoke \texttt{Delete()} on mappings associated with specific \ids, \eg, during maintenance operations such as PostgreSQL \texttt{VACUUM}, which removes dead tuples.

% The above partition approach incurs negligible runtime overhead.
% We embed a partition identifier in the high bits of each \id. For example, an \id with its high bits set to \texttt{TMP\_PARTITION\_NUM} indicates data in the temporary partition.
% This design adds only a constant number of bitwise operations to each \lut lookup, thus preserving the $O(1)$ access property.

\subsection{Exploiting Data Locality}
\label{s:locality}

Mismatches in data access patterns between the DBMS and \lut can cause \emph{major page faults} on the critical path.
For example, when the DBMS accesses an \id whose corresponding secret is not in memory, \sys must perform an extra I/O in the privacy zone, thereby increasing latency.
To reduce such latency, we make the \lut organization storage-layout-aware and proactively prefetch secrets likely to be accessed by the DBMS.
This is achieved by exploiting both temporal and spatial locality.

\heading{Exploiting temporal locality.}
% \sys has an inherent access-order constraint:
In \sys, the DBMS first retrieves the \id from storage, and then resolves it through the \lut to obtain the corresponding secret.
This lookup dependency creates a natural temporal alignment in access patterns between the DBMS and the \lut.
For instance, when a query accesses fields associated with $\text{FID}_1, \ldots, \text{FID}_n$ in sequence, the \lut performs the same sequence to retrieve $\text{secret}_1, \ldots, \text{secret}_n$.
This dependency also enables lock-efficient management for the \lut. Accesses to the same allocated \id do not introduce conflicts, because they are already coordinated by the DBMS's concurrency control. As a result, \sys synchronizes only during \id allocation, while reads and slot reuse need no additional locking.

\heading{Exploiting spatial locality.}
To further improve performance, \sys organizes the \lut into separate partitions based on the storage layout of the underlying database. The partitioning policy is customizable depending on the specific DBMS or storage engine.
For example, in PostgreSQL, each table is a separate heap file with unordered field placement and reclamation. \sys assigns one \lut partition per heap file to preserve field colocation.
In InnoDB, data resides in globally ordered fixed-size pages (\eg, 16 KB) with unordered fields inside each page. \sys assigns one partition per page to preserve page-level locality.
We encode the partition number in the high bits of \id to segment the \id space.
% Building on the embedding approach described in \autoref{s:kv}, we also encode the partition number in the high bits of each \id (excluding the highest bit) to segment the \id space.

\heading{Prefetching.}
A random lookup in \sys may involve two serial I/O operations: one for the DBMS to fetch \ids and another for the \lut to retrieve secrets. \sys reduces this overhead by maximizing I/O overlap.
Specifically, when the DBMS issues a disk read, \sys intercepts it and identifies the partition being accessed. Since \sys uses storage-layout-aware \lut partitioning, it proactively sends an RPC to the proxy specifying which partition will be accessed. The \lut then prefetches the corresponding partition concurrently with the DBMS's I/O operation.

\section{Ensuring Cross-domain Consistency}
\label{s:design:txn}

Modern CDBs rely on the underlying DBMS to provide ACID properties.
\sys must preserve the same guarantees. However, doing so in \sys is more subtle because transaction-relevant state now spans two isolated domains, requiring new mechanisms to keep these states consistent throughout the transaction lifecycle.

For cross-domain correctness, the key constraint is that the DBMS interacts with the privacy zone only through \ids; it cannot directly inspect the \lut state. Therefore, correctness reduces to a one-way invariant: \emph{every DBMS-maintained \id maps to a valid, up-to-date plaintext in the \lut}. In contrast, orphan plaintexts in the \lut are permissible because they are unreachable from DBMS-visible state. This asymmetry naturally aligns with \emph{external synchrony}~\cite{nightingale2008rethink}, a consistency model that distinguishes between internal system state and externally visible state. In \sys, the \lut constitutes the internal state, while the \ids maintained by the DBMS constitute the external state with respect to the privacy zone.

This observation pinpoints the moments that require cross-domain coordination: when the set of DBMS-visible \ids changes, or when such visibility must be preserved across failures. These cases include aborts, where old \ids become visible again after rollback; commits, where new \ids become durable; and recovery, where state is restored after failures. Accordingly, \sys introduces lightweight synchronization between the database and the \lut at transaction boundaries and during recovery. We describe these mechanisms next.

\heading{Handling transaction aborts.}
When a transaction aborts, rollback may make old \ids DBMS-visible again.
To preserve the invariant, \sys must keep the mappings for these old \ids until the DBMS itself determines that they are no longer visible.

\begin{figure}[t]
\centering
\includegraphics[width=\linewidth]{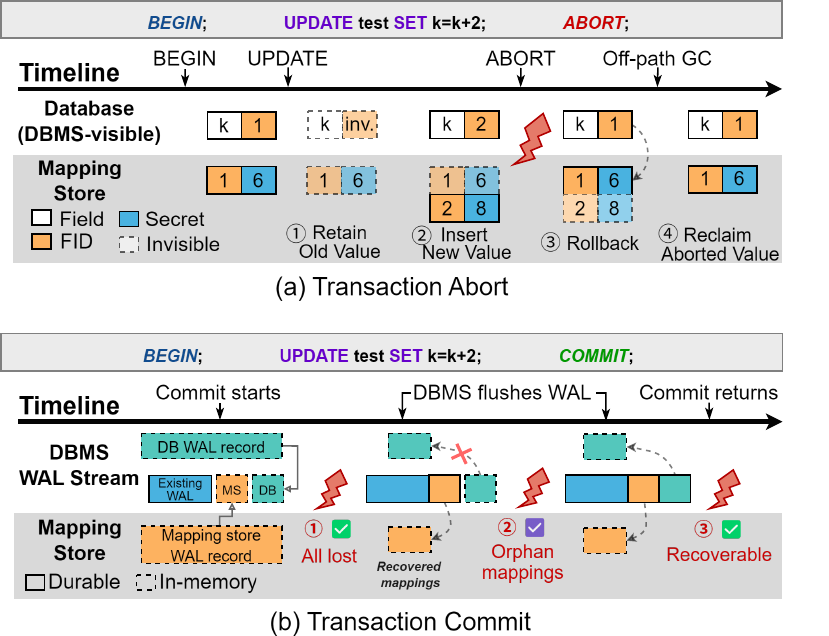}
\caption{Timeline of a transaction execution in \sys, illustrating the handling of (a) abort and (b) commit paths.}
\label{fig:txn-undo}
\vspace{-1.0em}
\end{figure}

To this end, \sys employs a multi-version concurrency control (MVCC)-like approach: stale values (\eg, logically deleted entries and pre-update versions) are retained at runtime and reclaimed only when the underlying DBMS runs garbage collection (\eg, PostgreSQL's off-path \texttt{VACUUM}). 
For instance, as illustrated in \autoref{fig:txn-undo}(a), when a transaction issues an update, \sys retains the original mapping entry intact (\ding{172}) and inserts the new value out-of-place (\ding{173}). If the transaction is aborted, the DBMS can access the original values as usual (\ding{174}). The aborted value is eventually reclaimed by the DBMS's off-path garbage collection process (\ding{175}).

\heading{Handling transaction commits.}
Commits must preserve the external-synchrony invariant across crashes. A privacy-zone crash may lose recently created secrets, while the DBMS can still recover the committed database state. After recovery, the DBMS may contain committed tuples with \ids whose corresponding secrets are no longer recoverable in the \lut.
To prevent this case, the commit protocol extends external synchrony to crash-recoverable state: if an \id is recoverable after a crash, its corresponding secret must also be. Since both PostgreSQL and the \lut recover committed state from their WALs, enforcing this invariant reduces to a commit-time WAL-ordering rule: for each transaction, mapping-store WAL updates must be durable before PostgreSQL persists its WAL.
A straightforward design would maintain a separate mapping-store WAL and synchronously flush it before PostgreSQL persists its own WAL. Although correct, this design is costly: under synchronous commit, the mapping-store flush becomes an additional serialized I/O on the critical path. This also forces \sys to track a separate global order for mapping-store WAL records, duplicating PostgreSQL's WAL ordering and increasing commit contention.

To avoid these costs, \sys co-designs mapping recovery with PostgreSQL's WAL recovery by embedding mapping-store WAL records into the PostgreSQL WAL stream. At commit time, \sys calls \texttt{SealLog()} to generate a block-granular encrypted mapping-store WAL record containing the newly exposed mappings from the transaction. It then adds this record to PostgreSQL's WAL stream before PostgreSQL emits the transaction commit record. Only after this insertion does PostgreSQL continue its normal commit path, placing the resulting WAL records in LSN order and flushing them using its native WAL machinery.
To further reduce commit-time overhead, \sys pipelines and parallelizes block-granular encryption: the privacy zone encrypts full blocks as they accumulate during transaction execution and handles multiple blocks in parallel. This protocol is therefore lightweight, adding only one RPC and residual block-granular encryption to the commit critical path.% while preserving PostgreSQL's native WAL ordering and optimizations such as group commit.

\heading{Recovering from system crashes.}
After a system crash, \sys must restore mappings for all \ids recovered by PostgreSQL. Recovery is driven by PostgreSQL's standard WAL replay. As PostgreSQL replays the WAL stream, \sys identifies embedded mapping-store WAL records and then calls \texttt{ReplayLog()} to reconstruct the corresponding mappings.

\autoref{fig:txn-undo}(b) illustrates three recovery outcomes.
In case~\ding{172}, the crash occurs before PostgreSQL flushes the WAL stream; both the mapping-store WAL record and the transaction commit record are lost, so no new \ids become visible after recovery.
In case~\ding{174}, the transaction commit record is durable; by order, the preceding mapping-store WAL record is also durable, so \sys can reconstruct mappings for all recovered DBMS-visible \ids.
Thus, both cases preserve the external-synchrony invariant.
Case~\ding{173} is the only intermediate case: the system crashes during WAL flushing, after the mapping-store WAL record is written back but before the transaction commit record is completely written to disk. WAL replay may reconstruct these mappings, but PostgreSQL will not expose their \ids in the recovered database state. This is safe because the recovered mappings do not overwrite pre-existing visible data: as described in the abort protocol, \sys only reuses \ids after the DBMS has garbage-collected their previous references. Therefore, the reconstructed mappings are merely orphans and do not constitute an inconsistent state. They remain unreachable from DBMS-visible state, preserving consistency; \sys can reclaim them later via a global scan.

\section{Implementation}
\label{s:impl}

We implemented \sys on top of PostgreSQL 15.5 in 6.1K lines of C/C++ code. \autoref{t:code-components} provides a detailed breakdown by component.
\sys runs atop CVMs (\ie, Intel TDX~\cite{intel-tdx} and ARM Secure EL2~\cite{armv8-manual, Li2021TwinVisor}) and deploys its integrity and privacy zones in isolated VM domains.
To reduce cross-domain RPC overhead, \sys adopts a context-switchless implementation~\cite{Antonopoulos2020AEv2, Arnautov2016SCONE, weisse2017hotcalls}: the integrity zone sends requests to secure shared memory, while the privacy zone polls the shared memory and dispatches requests for processing.
% To balance responsiveness against CPU contention from busy-waiting, \sys employs spin-then-park: workers busy-poll briefly, then sleep if no request arrives.
The integrity zone uses dm-integrity~\cite{dm-integrity} for corruption detection, while the privacy zone uses dm-crypt~\cite{dm-crypt} with HMAC-SHA256 for transparent I/O encryption and authentication.
The \lut is backed by \texttt{mmap}ed arrays for persistence. Temporal locality is exploited by an LRU cache combined with Linux sequential prefetching; spatial locality is improved by organizing permanent partitions at table granularity.
Within each partition, \sys assigns \ids by reusing safely reclaimed slots or appending new ones via an atomic monotonic counter.
Currently, each 64-bit \id encodes a configurable 16-bit partition prefix and a 48-bit offset. Users can also tune \id length and partition-prefix size according to expected data volume: shorter \ids reduce storage overhead and improve lookup efficiency.
An open-source version of \sys is available at \url{https://github.com/ISCAS-OSLab/ZENO}.

\begin{table}[t]
\centering
\caption{Lines of code for each component in \sys.} % including components inherited from HEDB
\vspace{-0.6em}
\scalebox{0.9}{
\begin{tabular}{l l r}
\toprule
\textbf{{\sys} Components} & \textbf{Language} & \textbf{LoC} \\
\midrule
Mapping Store & C++ & $1,074$ \\
WAL & C++ & $461$ \\
Commit Protocol (in PostgreSQL source) & C & $550$ \\
%\hline
\midrule
Proxy & C++ & $2,155$ \\
Operators (modified from HEDB source) & C++ & $1,708$ \\
Crypto Library & C/C++ & $110$ \\
\bottomrule
\end{tabular}
}
\label{t:code-components}
\end{table}

\section{Performance Evaluation}
\label{s:eval}

We evaluate \sys by answering the following questions:

\begin{myitemize2}
  \item RQ1: Does \sys deliver end-to-end performance improvements across platforms and workloads? (\autoref{s:eval-macro})
  \item RQ2: What are the sources of \sys{}'s performance gains and consistency overheads? (\autoref{s:eval-ablation})
  \item RQ3: How robust is \sys under different workloads and resource configurations? (\autoref{s:eval-micro})
\end{myitemize2}

\heading{Experimental setup.}
\label{s:eval-setup}
We evaluate \sys{} on two platforms.
The ARM platform runs on a HiSilicon Kunpeng-920 @ 2.6~GHz server with S-EL2 support, equipped with a 1~TB SSD and Ubuntu 22.04 LTS, hosting two Secure-EL2 VMs running Linux 6.8.0.
The x86 platform runs on an Intel Xeon Platinum 8581C @ 4.0~GHz server with TDX support, equipped with a 4~TB SSD and Ubuntu 24.04 LTS.

For both platforms, \sys{} and HEDB deploy the integrity and privacy zones in two isolated VM domains with a total resource budget of 32~vCPUs and 64~GB of memory.
Under this budget, we sweep several vCPU and memory splits between the two zones for each experiment and use the best-performing split for each system.
Both zones disable memory swapping.
To ensure resource fairness, baseline plaintext PostgreSQL is deployed in a virtual machine with 32~vCPUs and 64~GB of memory.
Across all systems, the benchmark client uses 16~vCPUs and 32~GB of memory.
To reduce cross-NUMA RPC latency and hypervisor scheduling noise, we pin all VMs to the same NUMA node when possible, or otherwise to nearby NUMA nodes with low NUMA distance.
% We pin the PostgreSQL client and server to disjoint CPU cores: the client runs on cores 0-15, while the PostgreSQL server runs on cores 16-31.
% We set the buffer pool size (\texttt{shared\_buffers}) to 25--40\% of the total available memory, following the PostgreSQL guideline~\cite{postgresql:runtime-config-resource}.
% Unless otherwise specified, we run each test \TODO{times} and report the geometric means.

For ARM S-EL2, the shared memory for inter-CVM RPCs is protected by the TrustZone hypervisor.
For x86 TDX, since no CVM vendors~\cite{intel-tdx,amd-sev} currently provide secure channels between CVMs (their shared memory remains plaintext), we exploit TD Partitioning~\cite{intel2023tdxpartitioning}, which allows multiple L2 VMs to run within one CVM. However, TD Partitioning is documented primarily for single-L2-VM deployments, offering little engineering guidance for multi-L2 setups. As such, we adopt a microkernel-inspired deployment: the privacy zone runs in the L1 VM and the integrity-zone DBMS (given its large TCB) runs in an L2 VM managed by the L1, with RPC shared memory controlled by the L1 to prevent malicious host introspection.
% For x86 TDX, we use TD Partitioning~\cite{intel2023tdxpartitioning} to protect inter-CVM RPCs because currently no CVM vendors provides secure channel between 2 CVMs (namely, shared memory between independent TDs are plaintext).
% Although TD Partitioning is designed to support multiple L2 VMs, it lacks adequate public materials for engineering guidance to build two-L2-CVM deployment. As such, we draw inspiration from microkernels, where operators in the privacy zone runs in the L1 VM, whereas DBMS in the integrity zone, due to its large TCB, runs in a L2 VM managed by the L1 VM.
% Two zones communicate through an L1-managed shared-memory region that is mapped into the L2 VM for RPC against host introspection.

\begin{figure}[t]
\centering
\begin{minipage}[t]{0.49\linewidth}
\centering
\includegraphics[width=\textwidth]{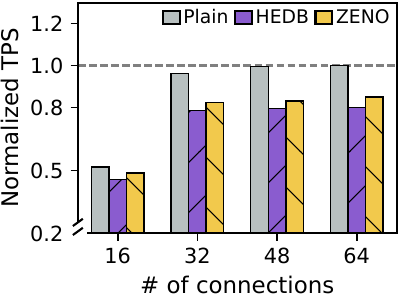}
\vspace{-0.5em}
\centerline{(a) ARM S-EL2}
\end{minipage}\hfill
\begin{minipage}[t]{0.49\linewidth}
\centering
\includegraphics[width=\textwidth]{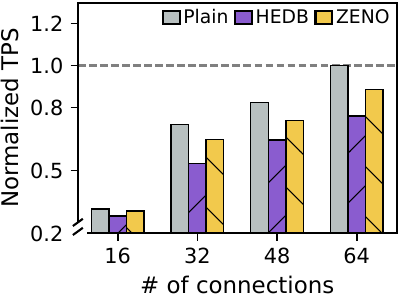}
\vspace{-0.5em}
\centerline{(b) x86 TDX}
\end{minipage}
\captionsetup{width=0.98\linewidth}
\caption{
TPC-C throughput with varying numbers of connections at $W=100$, normalized to the peak plaintext PostgreSQL TPS on each platform (higher is better).
\sys reduces HEDB's throughput loss by up to 49.8\% on ARM S-EL2 and 73.8\% on x86 TDX.
}
\label{fig:eval-tpcc-thread}
\vspace{-1.0em}
\end{figure}

\begin{figure*}[t]
\centering
\begin{minipage}[t]{0.49\linewidth}
  \centering
  \includegraphics[width=0.97\textwidth]{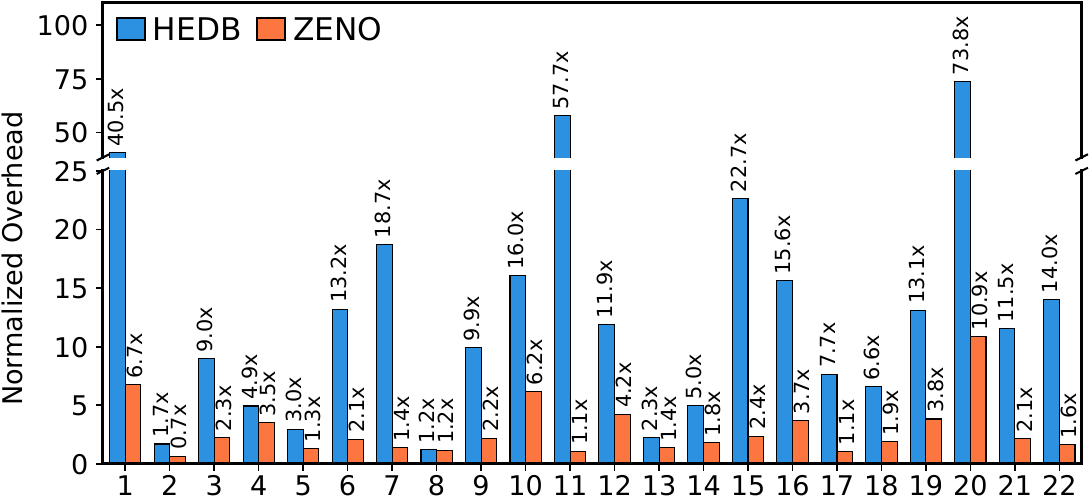}
  \vspace{-0.8em}
  \centerline{(a) ARM S-EL2}
\end{minipage}\hfill
\begin{minipage}[t]{0.49\linewidth}
  \centering
  \includegraphics[width=0.97\textwidth]{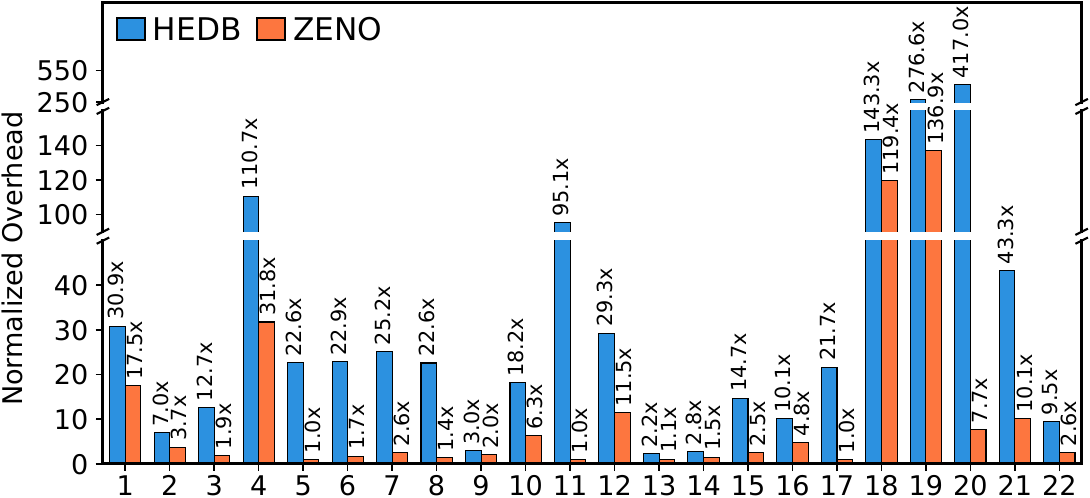}
  \vspace{-0.8em}
  \centerline{(b) x86 TDX}
\end{minipage}
\caption{
TPC-H execution time of HEDB and \sys{} normalized to plaintext PostgreSQL.
\sys{} substantially reduces HEDB's overhead, achieving speedups of up to $53.1\times$ on ARM S-EL2 and up to $94.7\times$ on x86 TDX.
}
\label{fig:tpch-overall}
\end{figure*}

\begin{figure*}[t]
\centering
\begin{minipage}[t]{0.49\linewidth}
  \centering
  \includegraphics[width=0.97\textwidth]{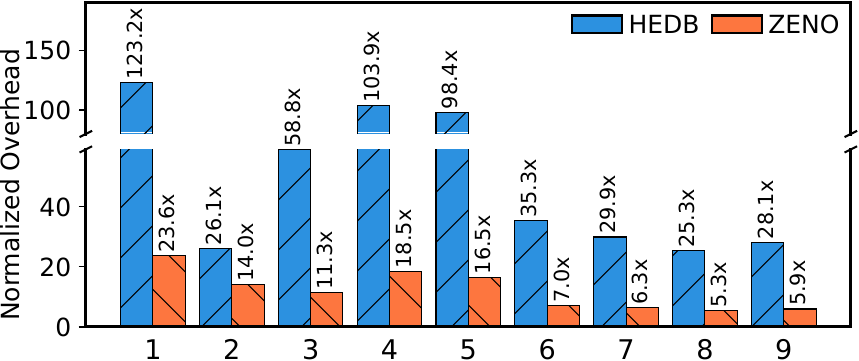}
  \vspace{-0.8em}
  \centerline{(a) ARM S-EL2}
\end{minipage}\hfill
\begin{minipage}[t]{0.49\linewidth}
  \centering
  \includegraphics[width=0.97\textwidth]{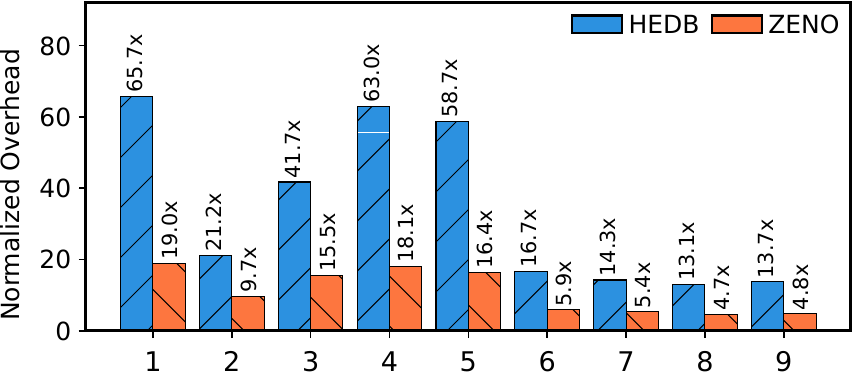}
  \vspace{-0.8em}
  \centerline{(b) x86 TDX}
\end{minipage}
\caption{{
Execution time of 9 industrial queries on ARM S-EL2 and x86 TDX, normalized to that of plaintext PostgreSQL.
\sys{} consistently outperforms HEDB, achieving up to 6.0$\times$ and 3.6$\times$ speedups on ARM S-EL2 and x86 TDX, respectively.
}}
\label{fig:industrial-overall}
\end{figure*}

\subsection{End-to-End Performance (RQ1)}
\label{s:eval-macro}

We use two standard benchmarks: TPC-C and TPC-H.
Since these standard benchmarks are not privacy-oriented, we further use a real-world workload from the industry.
For comparison, we include vanilla PostgreSQL as the plaintext baseline and the state-of-the-art HEDB.

\heading{TPC-C.}
We evaluate TPC-C throughput (transactions per second, or TPS) under varying connection counts using 100 warehouses, treating all non-ID columns as sensitive and placing the database on tmpfs.
After a 60-second warm-up, we measure for 300 seconds and normalize TPS to the peak plaintext PostgreSQL throughput across all connection counts.
As shown in \autoref{fig:eval-tpcc-thread}, \sys narrows the throughput gap between HEDB and plaintext PostgreSQL on both platforms. We report this gap as throughput loss relative to plaintext PostgreSQL at the same connection count.
On ARM S-EL2, \sys reduces HEDB's throughput loss by 18.1--49.8\%; on x86 TDX, \sys reduces HEDB's throughput loss by 51.5--73.8\%.
The two platforms exhibit different scaling trends due to platform-specific hardware and TEE-stack characteristics.

% The two platforms show different scaling trends: ARM S-EL2 plateaus after 32 connections, while x86 TDX continues to scale to higher connection counts. Together with our raw TPS measurements, this suggests that the higher per-core processing capability of our x86 TDX machine allows it to sustain more concurrent TPC-C workers before saturation.

% On ARM S-EL2, \sys{} improves throughput from HEDB's 70.8--97.1\% of plaintext to 81.4--98.6\%, yielding an 11.8\% average throughput gain over HEDB. The benefit is consistent on x86 TDX, where \sys{} raises throughput from 47.4--98.5\% to 57.0--98.8\% of plaintext, with a 17.0\% average gain.
% All three systems scale nearly linearly at low concurrency, after which scaling tapers off as CPU cores become saturated.
% The plaintext PostgreSQL baseline reaches this saturation point later because it has more CPU resources.
% Both platforms exhibit similar scaling trends.
% Throughput stops scaling and eventually drops at high concurrency due to increased contention on warehouse accesses inside the database.
% HEDB and \sys{} perform similarly at low concurrency, but \sys{} gains a larger advantage under high concurrency because its lighter privacy-zone processing allows more CPU resources to be assigned to the integrity zone.
% In our best-performing configurations, \sys{} uses an average integrity/privacy vCPU split of 29:3, compared with 27:5 for HEDB, which helps alleviate the PostgreSQL-side bottleneck.

\heading{TPC-H.}
We use TPC-H (scale factor = 3), a standard OLAP benchmark with large joins, group-bys, aggregations, and arithmetic operations.
We make all non-key columns sensitive, as TPC-H does not define security attributes.
To evaluate performance under large-dataset execution, we use a 6 GB memory budget. PostgreSQL and HEDB each use the full 6 GB as DBMS-side memory, while \sys splits the same total budget between the integrity and privacy zones, with 4.5 GB and 1.5 GB, respectively.
% Both \sys and HEDB enable batching (with a batch size of 256 fields), which aggregates multiple operations into a single RPC instead of iteratively.
% We also report results with batching disabled.
Results are averaged over 5 runs per query after one warm-up run.
\autoref{fig:tpch-overall} shows the slowdown of HEDB and \sys relative to plaintext PostgreSQL.
HEDB incurs an average slowdown of 10.0$\times$ on ARM S-EL2 and 23.8$\times$ on x86 TDX, driven by excessive cryptographic operations and I/O amplification due to ciphertext expansion.
In contrast, \sys restricts the average slowdown to 2.3$\times$ on ARM S-EL2 and 4.5$\times$ on x86 TDX, outperforming HEDB by 4.4$\times$ and 5.3$\times$, respectively.
The performance discrepancy is primarily attributable to the Intel TDX implementation, which relies on bounce buffers (\ie, Linux SWIOTLB) for I/O~\cite{li2023bifrost}. Consequently, I/O-intensive queries experience higher overhead on x86 TDX compared to ARM S-EL2.
% Across both platforms, disabling batching degrades performance for both systems, since aggregation operators like \texttt{SUM/AVG/MIN/MAX} can no longer process data in bulk.
% Even without batching, \sys remains faster than HEDB on most queries.
% Interestingly, \sys slightly outperforms plaintext PostgreSQL on QX on ARM and QX on x86 TDX.
% This occurs because \sys stores only fixed-size \ids, enabling PostgreSQL's query planner to apply more aggressive and efficient physical optimizations than when handling variable-length types (\eg, \texttt{VARCHAR}, \texttt{DECIMAL}) in plaintext.

\begin{table}[t]
    \centering
    \caption{Overall storage usage (in GB) across three workloads. \sys reduces storage cost by up to 52.8\% compared to HEDB.}
    \label{tbl:eval-storage-cost}
    \vspace{-0.5em}
    \resizebox{\linewidth}{!}{%
    \begin{tabular}{c|lll}
        \toprule
        \textbf{Database}
        & \textbf{TPC-C (W=100)}
        & \textbf{TPC-H (SF=3)}
        & \textbf{Industrial} \\
        \midrule
        Plaintext & 10.3 & 5.2 & 1.5 \\
        HEDB      & 22.1 (2.1$\times$) & 17.8 (3.4$\times$) & 3.3 (2.2$\times$) \\
        \sys      & 13.5 (1.3$\times$) & 8.4 (1.6$\times$) & 1.9 (1.3$\times$) \\
        \bottomrule
    \end{tabular}
    }
\vspace*{-1em}
\end{table}

\heading{Industrial workload.}
We evaluate \sys using 9 real-world complex queries under strict confidentiality constraints. Schemas are anonymized prior to evaluation.
We generate synthetic data matching the anonymized schemas at 1 million rows per table, with all columns treated as sensitive.
\autoref{fig:industrial-overall} reports the end-to-end execution time normalized to plaintext PostgreSQL.
\sys outperforms HEDB on both ARM S-EL2 and x86 TDX.
% On ARM S-EL2, HEDB incurs a 55.4$\times$ average slowdown, whereas \sys reduces the slowdown to 12.5$\times$, delivering up to 6.0$\times$ speedup over HEDB.
% On x86 TDX, HEDB incurs a 34.2$\times$ average slowdown, whereas \sys reduces the slowdown to 11.0$\times$, delivering up to 3.6$\times$ speedup over HEDB.
On ARM S-EL2, \sys reduces HEDB's 48.4$\times$ average slowdown to 10.5$\times$ (up to 6.0$\times$ speedup);
on x86 TDX, it reduces 27.6$\times$ to 9.4$\times$ (up to 3.6$\times$ speedup). 
These results confirm \sys's performance gains on real-world industrial workloads.

\heading{Storage consumption.}
\label{s:eval-storage}
As shown in \autoref{tbl:eval-storage-cost}, \sys reduces storage consumption by 38.9--52.8\% compared to HEDB across all evaluated workloads. This is because \sys adds an 8-byte \id per value, while existing modern CDBs attach 28 bytes of cryptographic metadata per field.
% In addition, HEDB requires padding each sensitive field to an upper bound to hide sizes (if necessary), which further amplifies I/O and memory pressure.
% In contrast, \sys by design hides the size information as it replaces all fields with constant-size \ids.
% \TODO{(256 B, 128 B, and 48 B for the three workloads, respectively)}.

\subsection{Breakdown and Overhead Analysis (RQ2)}
\label{s:eval-ablation}

To address RQ2, we use TPC-H and TPC-C to isolate different performance factors.
TPC-H is ideal for decomposing \sys{}'s gains; its heavy aggregations and large data volumes expose the specific overheads of cryptographic computation and ciphertext expansion. For aspects not covered by TPC-H, specifically the commit protocol and crash recovery, we use TPC-C to measure their overhead.

% \TODO{REVIEW: explain why we use ARM for the following section}
Unless otherwise stated, the following experiments use our ARM S-EL2 VM setup.
These studies address platform-independent design questions. Fixing one primary platform avoids conflating \sys{}'s design effects with platform-specific factors such as I/O-path behaviors (\eg, bounce buffers), TEE virtualization overheads, and scheduling.
% Besides, ARM serves as our fully calibrated evaluation environment for the complete benchmark suite, including resource isolation, storage configuration, and baseline comparisons.

\begin{figure}[t]
\centering
\includegraphics[width=\linewidth]{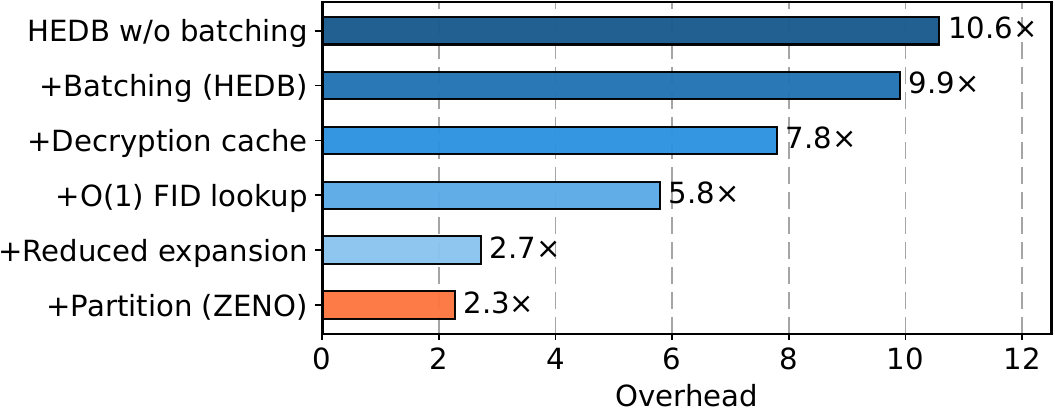}
\caption{{
Factor analysis of performance gains across different techniques using TPC-H. Overhead is the execution time normalized to the plaintext baseline.
}}
\label{fig:tpch-ablation}
\vspace*{-0.5em}
\end{figure}

\heading{Factor analysis.}
We perform a cumulative factor analysis on TPC-H to characterize the sources of \sys{}'s performance gains.
Starting from unoptimized HEDB, we incrementally enable HEDB's optimization and then introduce \sys's optimization techniques one by one.
\autoref{fig:tpch-ablation} reports the geometric mean execution time normalized to plaintext PostgreSQL.
Below, all reported reductions and speedups are relative to the preceding variant.

\emph{+Batching} groups aggregation inputs for operators such as \texttt{SUM} and \texttt{AVG}.
Instead of invoking the privacy zone for every intermediate accumulation step, the integrity zone sends a batch of fields per RPC (batch size = 256), reducing RPC frequency and intermediate cryptographic operations.
Batching reduces the execution time by 5.8\% and speeds up individual queries by up to 1.6$\times$.
% The improvement is limited because each input item still requires decryption and the expanded ciphertext representation remains unchanged.

\emph{+Decryption cache} represents an optimized version of the ciphertext-to-plaintext cache discussed in \autoref{s:motivation}.
It caches decryption results in the privacy zone, using \texttt{boost\_unordered\_flat\_map}~\cite{boost_unordered} with \texttt{XXH3\_64bits}~\cite{xxhash} and a 512~MB cache shared across SQL statements.
The decryption cache further reduces the execution time by 21.8\% and speeds up individual queries by up to 2.4$\times$.
To understand why the gains remain modest, we further instrument per-operation costs under the evaluated workload.
A hash-table lookup and an insertion cost 334.9 and 2649.4 cycles on average, respectively, compared with 113.8 and 316.0 cycles for \sys's direct-indexing \texttt{Get()} and \texttt{Put()}.
Although avoiding repeated decryptions is beneficial, the improvement is bounded by two factors: the relatively high cost of hash-table lookup/insertion operations over ciphertext keys, and the overhead of ciphertext expansion, which remains unchanged.

\emph{+O(1) \id lookup} replaces each ciphertext with a 28-byte padded \id, which is resolved through direct indexing in the mapping store.
Together, the padded \id and plaintext value match HEDB's per-field ciphertext footprint, separating the benefit of direct indexing from that of reduced data expansion.
O(1) \id lookup reduces execution time by 25.7\% and speeds up individual queries by up to 24.7$\times$, confirming that direct indexing substantially lowers privacy-zone processing overhead.

\emph{+Reduced expansion} switches to compact 8-byte \ids, removing field-level expansion and reducing the data footprint of TPC-H's large, I/O-bound queries. By reducing I/O and memory movement, this step lowers execution time by 53.0\% and speeds up individual queries by up to 11.5$\times$.

Finally, \emph{+Partition} enables \sys{}'s partition-based mapping-store layout. By organizing partitions according to the DBMS storage layout, \sys{} improves spatial locality between DBMS-visible \ids and privacy-zone plaintext values. This optimization further reduces the execution time by 16.6\% and speeds up individual queries by up to 6.4$\times$.

Overall, the factor analysis confirms the combined benefits of direct \id lookup, compact field representation, and locality-aware partitioning.

\begin{figure*}[t]
    \centering
    \begin{minipage}[t]{0.33\linewidth}
        \centering
        \includegraphics[height=0.45\linewidth, width=0.92\textwidth]{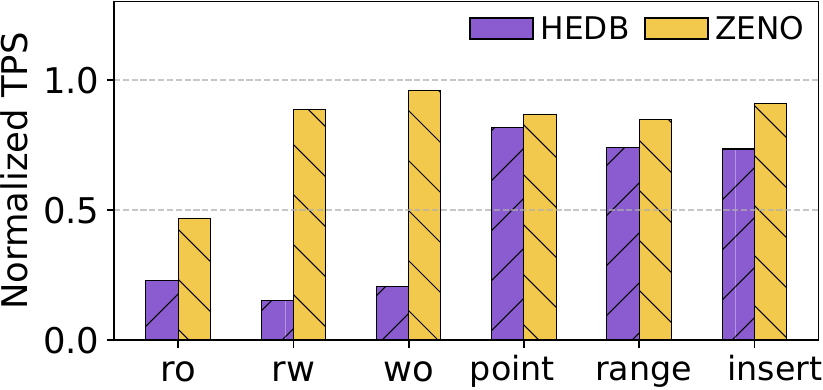}
        \makebox[0.52\textwidth]{\hspace{1em}(a) Transaction Modes (Zipfian, factor=0.8)}
        \vspace{-0.5em}
    \end{minipage}
    \begin{minipage}[t]{0.33\linewidth}
        \centering
        \includegraphics[height=0.45\linewidth, width=0.92\textwidth]{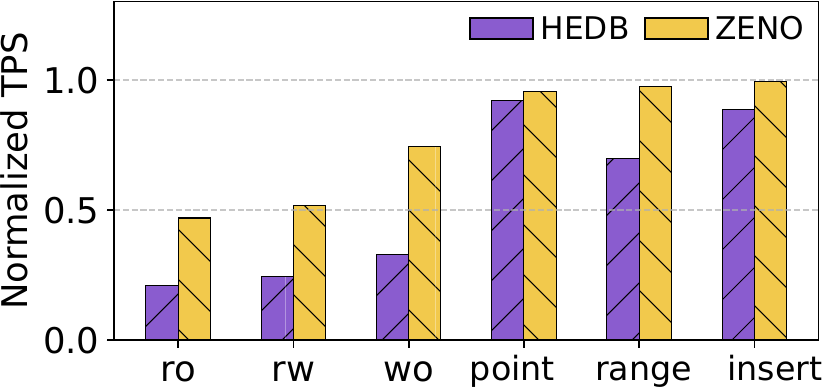}
        \makebox[0.52\textwidth]{\hspace{2em}(b) Transaction Modes (Uniform)}
        \vspace{-0.5em}
    \end{minipage}
    \begin{minipage}[t]{0.33\linewidth}
        \centering
        \includegraphics[height=0.45\linewidth, width=0.92\textwidth]{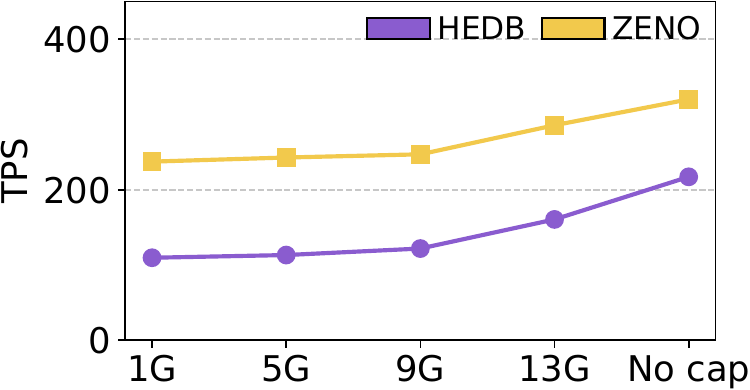}
        \makebox[0.52\textwidth]{\hspace{2em}(c) Total Memory Size}
        \vspace{-0.5em}
    \end{minipage}

    \caption{{Microbenchmark performance under different configurations. \sys consistently outperforms HEDB across all settings.} }
    \label{fig:eval-micro-pattern}
    \vspace*{-0.5em}
\end{figure*}

\heading{Commit overhead.}
To capture the runtime overhead of \sys{}'s commit protocol, we place the database on persistent storage and enable \texttt{synchronous\_commit}, forcing each transaction commit to wait for WAL durability.
We use TPC-C with $W=100$ and 16 connections for this experiment, and compare \sys{} with and without embedded mapping-store WAL records.
Enabling mapping-store WAL incurs a modest 2.6\% reduction in TPS.
To further understand this overhead, we instrument the commit path and directly measure the per-commit cost introduced by \sys{}'s commit protocol.
% This penalty is driven by two commit-path overheads: persisting extra WAL bytes and inserting mapping-store records (along with residual block-granular encryption) into the PostgreSQL WAL stream.
An invocation producing a 997-byte encrypted payload adds only $18.3\,\mu\mathrm{s}$ to the overall commit latency.

\heading{Recovery overhead.}
During the initial data loading phase of TPC-C (configured with 8 warehouses and 8 threads), we abruptly terminate the process to simulate a crash, leaving 2.6 GB of WAL for recovery.
After restart, PostgreSQL replays the WAL stream that contains both database records and embedded mapping-store records.
The complete WAL replay takes 26.3 seconds, with mapping-store record replay accounting for 4.1 seconds (15.6\%) of the total time.

\subsection{Workload and Resource Sensitivity (RQ3)}
\label{s:eval-micro}

The performance gain of \sys depends on both workload characteristics and system configuration.
To characterize the impact of these factors, we use Sysbench~\cite{oracle2021sysbench} to issue transactional microbenchmarks with configurable read/write ratios and access patterns.
The benchmark runs on a synthetic dataset consisting of 32 tables with 1M rows each, occupying approximately 7 GB in plaintext PostgreSQL, 10 GB in \sys, and 12 GB in HEDB.
To control for other variables, we conduct all experiments entirely in memory, with the exception of the memory-sensitivity analysis.
% When experimenting with memory sensitivity, we use cgroups to adjust the total available memory (the sum of the integrity and privacy zones) to match different sizes of the dataset.
% Unless otherwise specified, each transaction randomly selects a table to access, using a uniform distribution over all tables.
% \sys's performance gain depends on various external factors (\eg, workload characteristics) and internal factors (\eg, system configuration). In this section, we characterize how these factors influence the performance of \sys.
% Each record is 188 bytes in size, consisting of an unencrypted primary key and three encrypted fields: one 32-bit integer and two strings.
Each experiment includes a 120-second warm-up followed by a 600-second measurement, using 16 connections by default.

\heading{Transaction sensitivity.}
\autoref{fig:eval-micro-pattern}(a) presents normalized throughput under Zipfian data distribution (factor = 0.8) across six operation modes: read-only, read-write, write-only, point-select, range-select and insert-only.
Overall, \sys achieves speedups of 1.1--5.8$\times$ over HEDB and delivers 46.6--95.0\% of plaintext throughput.
% The write-only mode involves \texttt{INSERT/UPDATE/DELETE} queries.
In write-only mode, \sys is 4.6$\times$ faster than HEDB. This gap arises from frequent \texttt{UPDATE} queries such as ``\texttt{UPDATE test SET k=k+1;}'', where \sys eliminates the cryptographic overhead in HEDB's decrypt-compute-re-encrypt cycle.
% The read-heavy mode involves intensive secret computation, and thus can benefit from the \cfmaps.

\heading{Skew sensitivity.}
We also evaluate a random access pattern generated from a uniform distribution, which exhibits relatively poor locality.
As illustrated in \autoref{fig:eval-micro-pattern}(b), even under a non-skewed access pattern that yields frequent CPU cache and TLB misses, \sys still outperforms HEDB by up to 2.3$\times$ across all six modes, thanks to its \cfmaps.
% We then adjust the Zipfian factor from 0.6 to 0.99 in read-write mode.
% \autoref{fig:eval-micro-pattern}(c) shows that all systems profit from higher skew as hot keys concentrate in cache, yet \sys outperforms HEDB and remains close to the plaintext baseline. These results show that \sys's performance advantage holds regardless of access locality.
% : \sys scales from 1.9 KTPS to 2.4 KTPS, while HEDB crawls from 0.24 KTPS to 0.74 KTPS.

\heading{Memory sensitivity.}
Both HEDB and \sys introduce additional metadata overhead (cryptographic metadata and \ids, respectively) over PostgreSQL, making them more sensitive to available memory size.
% % major page fault ~= cache miss. But ai tell me that page fault is a more common description between memory and disk
% Memory constraints impact HEDB and \sys differently.
% In contrast, HEDB suffers from much more severe storage expansion (as shown in \autoref{tbl:eval-storage-cost}).
% % On one hand, \sys's overall performance is related to the average mapping lookup latency, which will be increased because limited memory will trigger very frequent major page faults on the critical path.
% % HEDB, on the other hand, suffers from much more severe storage expansion (\autoref{tbl:eval-storage-cost}), which makes I/O significantly more frequent than in plaintext PostgreSQL under the same memory budget.
% \sys's overall performance is directly related to the average mapping lookup latency; limited memory will trigger frequent major page faults on the critical path.
To quantify this sensitivity, we measure their performance under different total memory sizes. We use a read-write workload with 8 threads and a random access pattern. 
% , representing a typical transactional workload with poor locality.
As illustrated in \autoref{fig:eval-micro-pattern}(c), \sys outperforms HEDB, with an average speedup of 1.9$\times$ and a 2.2$\times$ speedup even when the total available memory is limited to 1 GB (10\% of the dataset for \sys).
This is because \sys has a smaller storage footprint than HEDB and preserves locality between the \lut and the database.

\section{Security Analysis}
\label{sec:security}

% This section analyzes \sys's security guarantees. We first establish our security objectives (\autoref{sec:security-objectives}), then demonstrate how \sys achieves these objectives against both active attacks (\autoref{sec:active-attacks}) and passive attacks (\autoref{sec:passive-attacks}). Finally, we discuss residual side-channel risks and mitigation strategies (\autoref{sec:side-channels}).
This section analyzes \sys's security guarantees.
Following prior modern CDBs~\cite{Vinayagamurthy2019StealthDB, Antonopoulos2020AEv2, Guo2021GaussDB, Wang2022Operon, Li2023HEDB}, we consider two adversary classes:
active adversaries can arbitrarily tamper with storage and modify untrusted memory, and
passive adversaries (honest-but-curious) do not modify data but attempt to infer sensitive information through observation.

\heading{Defense against active attacks.}
\label{sec:active-attacks}
% \heading{Data confidentiality and freshness.}
\sys ensures that sensitive data remains confidential and fresh.
First, TEE hardware provides memory encryption and isolation, preventing adversaries from accessing plaintext data during computation.
Second, since \ids are completely opaque to users, a malicious user cannot access others' private data by crafting an \id.
Last, \sys secures untrusted storage against tampering across both zones.
% \sys splits each field into two components---an \id stored in the integrity zone and the corresponding plaintext secret in the privacy zone---both protected within their respective TEEs during processing. Third, 
In the privacy zone, \sys uses dm-crypt~\cite{dm-crypt} with HMAC-SHA256 to provide block-level authenticated encryption and detect storage corruption.
In the integrity zone, \sys enforces full-disk integrity \cite{dm-integrity} to guard against adversaries who may attempt to swap \ids across records to gain unauthorized access to a victim's secret.
% Instead of heavy per-field MACs like prior modern CDBs, \sys authenticates data context via the partition technique (\autoref{s:locality}). During retrieval, the DBMS verifies an \id's partition prefix against the expected table or tenant; any cross-partition swap thus triggers a mismatch and an immediate abort.
Both zones exploit a counter-based freshness mechanism~\cite{arm-rollback-protection} against rollback attacks~\cite{Angel2023Nimble,chu2026rollbaccine}.

% \heading{Execution integrity.}
\sys ensures faithful query execution through two mechanisms. First, remote attestation allows users to verify that legitimate \sys code is running in the TEEs before uploading data.
Second, in our implementation based on HEDB, \sys places both the DBMS (in the integrity zone) and expression operators (in the privacy zone) within trusted domains, with secure channels between TEEs that prevent message injection, replay, man-in-the-middle, and smuggle attacks.

\heading{Defense against passive attacks.}
\label{sec:passive-attacks}
% Passive attackers attempt to infer sensitive information by observing \sys's external behavior.
% We analyze what information \sys protects and what it intentionally reveals.
% \heading{Semantic security.}
The primary concern is whether an adversary observing \ids can learn anything about the corresponding plaintext secrets. We prove that \sys achieves indistinguishability via a simple observation: \id allocation is \emph{stateful but plaintext-independent}.
The allocator maintains internal state that identifies the next available \id and assigns \ids deterministically from this state, independent of the plaintext $m$. Consequently, after $n$ prior \texttt{Put()} calls, either plaintext $m_0$ or $m_1$, when submitted as the $(n+1)$-th item, receives the exact same \id, so the adversary's view of the \id sequence is indistinguishable between the two cases. As a result, \sys's \cfmaps provide confidentiality comparable to \emph{semantic security}~\cite{goldwasser1982probabilistic}: no information about plaintexts is revealed (beyond leakage from other channels, discussed later).

% \sys not only provides \emph{confidentiality}---adversaries cannot derive any information about plaintexts from \ids---but also 
Additionally, because \ids are assigned independently of plaintext values, the same plaintext may be represented by multiple distinct \ids, mirroring the behavior of non-deterministic encryption schemes like AES-GCM.

% \TODO{PLEASE REVIEW HERE: Revise leakage analysis to explicitly include \id-related metadata in the leakage profile, then add a separate paragraph to interpret this metadata in context, analogous to the later side-channel discussion.}

\heading{Leakage analysis.}
% \sys protects plaintext content but leaks certain metadata to enable functional query processing.
% The primary leakage is ordering relationships, revealed by comparison operations required for indexing and efficient execution—a fundamental trade-off inherent to modern CDBs~\cite{Antonopoulos2020AEv2,Guo2021GaussDB,Wang2022Operon,Li2023HEDB}.
% In addition, \sys leaks total storage volume per user, but provides stronger privacy for size information of each individual field, because it replaces variable-length fields with constant-size \ids.
% Finally, \sys does not protect against I/O access patterns and timing (at what time users query and how long they take) leakage, as such defenses would conflict with performance goals.
%
% Like HEDB, \sys provides \emph{bounded leakage} for maintainability: only queries that users explicitly consent to replay (for debugging purposes) expose their metadata to DBAs. During normal operation, the DBMS runs within the integrity zone TEE, limiting DBA visibility.
%
While \sys protects plaintext contents, practical query functionality requires revealing certain metadata to the untrusted infrastructure. We characterize \sys's leakage profile using the definition of $\mathcal{L}$-security~\cite{Poddar2016Arx,Vinayagamurthy2019StealthDB}, which models security as a leakage function $\mathcal{L}$ that precisely specifies what information an adversary learns.
We define \sys's $\mathcal{L} = (\mathcal{L}_{init}, \mathcal{L}_{query}, \mathcal{L}_{update})$ as follows:
\begin{myitemize}
\item $\mathcal{L}_{init}(D)$: During database initialization with dataset $D$, the adversary observes the database storage layout, including its total volume and the number and positions of protected fields. Since these fields are replaced by constant-size \ids, individual field sizes remain hidden.
\item $\mathcal{L}_{query}(q, D)$: For query $q$ over $D$, the adversary learns:
\begin{myenumerate2}
\item \emph{I/O access pattern} $ap(q)$: which storage blocks in the DBMS and \lut are accessed during query execution.
% In \sys, \ids index \lut entries organized into size-class buckets stored as contiguous arrays.
% Observing accesses to a \lut block reveals its entry density (the number of \id-indexed entries it contains).
% , identifying the corresponding size-class bucket and exposing precise mapping-store locations of the accessed fields.
\item \emph{Search pattern} $sp(q)$: whether two queries access overlapping record sets, revealing query repetition.
\item \emph{Timing} $t(q)$: when and how long the query executes.
\item \emph{Result volume} $|R_q|$: the number of records returned or affected by the query.
\item \emph{Comparison results}: the boolean outcomes (\eg, ``$>$'') necessary for indexing and filtering, a common trade-off for many enterprise modern CDBs~\cite{Antonopoulos2020AEv2,Guo2021GaussDB,Wang2022Operon,Li2023HEDB}. % for predicates involving protected columns
\end{myenumerate2}
\item $\mathcal{L}_{update}(op, D)$: For an update operation $op$ (\ie, \texttt{INSERT}, \texttt{UPDATE}, \texttt{DELETE}), the adversary learns which storage blocks are modified and the operation type, but not the plaintext values being written or deleted. Since \ids are allocated monotonically by default, the adversary may also infer the order in which protected fields are assigned \ids.
\end{myitemize}
This leakage profile provides \emph{$\mathcal{L}$-semantic security}: an adversary observing \sys's execution learns nothing beyond $\mathcal{L}$ about the underlying plaintext data.
% Formally, for any two databases $D_0, D_1$ and query sequences $Q_0, Q_1$ such that $\mathcal{L}(D_0, Q_0) = \mathcal{L}(D_1, Q_1)$, the adversary's view is computationally indistinguishable.

\heading{\id metadata discussion.}
Like ciphertexts in HEDB, \ids carry observable metadata about protected fields.
First, \ids expose positional information (via embedded partition numbers), analogous to how HEDB leaks ciphertext position through untrusted storage layout. % Critically, storage layout is thus independent of plaintext content.
Second, \ids reveal allocation order, as they are allocated monotonically by default. This information is also observable in HEDB: newly generated ciphertexts are eventually persisted to untrusted storage, and their appearance in logs or batch flushes can reveal generation order, at least at batch granularity. To coarsen this metadata, \sys could pre-allocate \ids in batches and assign them via random permutation within each batch, which balances allocation-order confidentiality, allocation efficiency, and spatial locality.

\heading{Side-channel discussion.}
\label{sec:side-channels}
\sys's stateful privacy zone retains plaintext data in TEE memory for extended periods (until eviction to storage). This residency could increase the attack window for digital side-channels (\eg, cache timing or memory access patterns). However, modern CDBs already face this risk: encryption keys reside permanently in TEE memory; key compromise would breach all sensitive data.
% \sys's stateful privacy zone retains plaintext data in TEE memory until eviction, extending the attack window for digital side-channels (\eg, cache timing or memory access patterns). However, modern CDBs already face an analogous risk: encryption keys reside permanently in TEE memory, so key compromise would breach all sensitive data.
Mitigations such as strict resource isolation~\cite{zhou2023coreslicing, castes2024coregapping} apply equally to both residual plaintext and long-lived keys.

% \TODO{We leave efficient and scalable obliviousness of data access (\eg, \cite{Mavrogiannakis2025OBLIVIATOR}) for \lut as future work.}
% Compared with key-only residency, however, \sys may expose a larger side-channel footprint. We rely on orthogonal TEE hardening techniques to reduce such implementation-level leakage. For one important class, controlled-channel attacks~\cite{xu2015controlledchannel},
% % \sys mitigates controlled- or side-channel attacks by leveraging modern TEE isolation features.
% \sys utilizes ARM Secure EL2~\cite{armv8-manual} that fully isolates TEEs and their page tables from untrusted privileged software.
% Importantly, \sys's design is generalizable to other TEEs offering equivalent guarantees, including Intel TDX~\cite{intel-tdx}, AMD SEV-SNP~\cite{amd-sev}, IBM PEF~\cite{Hunt2021PEF}, etc.
% Physical attacks that infer DRAM access patterns remain prohibitively expensive~\cite{Lee2020OffChip}.

\section{Related Work}
\label{s:rlwork}

\heading{Confidential databases (CDBs).}
% \TODO{REVIEW: generality to other Type-II CDB}
To address rising privacy concerns, CDBs on untrusted clouds have long been a goal~\cite{Maheshwari2000TDB}.
TrustedDB~\cite{Bajaj2011TrustedDB}, EnclaveDB~\cite{Priebe2018EnclaveDB}, DBStore~\cite{Ribeiro2018DBStore}, and SecuDB~\cite{yang2024secudb} shield an entire DBMS in TEEs, providing strong security but limiting maintainability.
Modern CDBs, such as AlwaysEncrypted~\cite{Antonopoulos2020AEv2}, GaussDB~\cite{Guo2021GaussDB}, Operon~\cite{Wang2022Operon}, and HEDB~\cite{Li2023HEDB}, prioritize maintainability by keeping most DBMS logic outside TEEs, but incur severe overhead from per-field cryptography.
Other CDB systems take orthogonal approaches: fully cryptography-based databases~\cite{Popa2011CryptDB,Poddar2016Arx,Tu2013Monomi, bian2023he3db, ren2022heda, bian2025engorgio, zhang2024arcedb, zhao2025hermes} require no trusted hardware, and oblivious databases~\cite{mishra2018oblix,crooks2018obladi,zhu2025compass} hide access patterns.

% \sys targets the latter---an emerging, industry-favored type to deliver far higher efficiency without compromising confidentiality or maintainability.
% Motivated by the growing industrial adoption of split-architecture CDBs, \sys improves their efficiency through crypto-free mappings: ciphertexts can be replaced with \ids, and trusted operators can replace en/decryption with \texttt{put/get} operations.
% Security against active DBAs, however, requires interface security. Among these systems, HEDB provides interface security by isolating DBAs from operator interfaces and authenticating valid user-query invocations. Systems such as AlwaysEncrypted, GaussDB, and Operon can adopt \sys's mapping abstraction as a performance optimization, but would need HEDB-style interface security to achieve HEDB-level security under our active-DBA threat model.
Motivated by the growing industrial adoption of split-architecture CDBs, \sys improves their efficiency through \cfmaps. Full resistance to active DBAs in this setting further requires defending against smuggle attacks, which HEDB addresses by isolating DBAs from operator interfaces.
Systems such as AlwaysEncrypted, GaussDB, and Operon could adopt \sys's mapping abstraction as a drop-in performance optimization, yet would still need HEDB-style isolation to achieve equivalent security guarantees.

\heading{Indirection-protection separation.}
Separating indirection from protection is a classic systems principle.
Capability systems~\cite{dennis1980capability, levy2014capability} pioneered the use of unforgeable tokens (capabilities) as references to resources protected by OS isolation.
% , relying on OS isolation rather than cryptography for protection.
Similarly, OSes have long used opaque handles~\cite{ritchie1974unix, accetta1986mach}: file descriptors in Unix are integers mapped by the kernel to internal objects.
% , enabling user processes to manipulate resources without direct access.
% In both cases, indirection need not carry protection; security comes from address-space isolation, keeping references simple and fast.
\sys shares this insight: data-independent \ids serve as capabilities protected by TEE isolation, removing cryptographic overhead for every reference.

\heading{Applicability beyond CDBs.}
Compartmentalization-based confidential analytics systems~\cite{zheng2017opaque, Kim2025Laputa} encrypt data at TEE boundaries among operators, incurring significant cryptographic overhead.
These systems may benefit from the abstraction of \cfmaps for better efficiency.
% such as Spark-based (e.g., Opaque~\cite{zheng2017opaque}, Laputa~\cite{Kim2025Laputa}) and Hadoop-based (e.g., VC3~\cite{schuster2015vc3}, Civet~\cite{Tsai2020Civet}),
% Sensitive big data analysis such as Opaque~\cite{zheng2017opaque}, VC3~\cite{schuster2015vc3}, and Ryoan~\cite{hunt2016ryoan} have reported overhead due to frequent cryptographic operations across TEE trust boundaries.

\heading{TEE optimizations.}
Several techniques have been proposed to optimize TEE-based applications, such as context-switchless calls~\cite{Antonopoulos2020AEv2, Arnautov2016SCONE, weisse2017hotcalls}, userspace I/O~\cite{bailleu2019speicher, Thalheim2021rktio}, and other optimizations~\cite{kim2019shieldstore, Li2023HEDB}.

\section{Conclusion}

% While intuitive in hindsight, the idea that cryptographic mappings are unnecessary contradicts conventional wisdom in CDB design.
This paper challenges a core assumption in modern CDB design: mappings between untrusted and trusted domains must be cryptographic.
% % For decades, the CDB community has treated encryption as the sole mechanism for protecting data references outside TEEs.
% \sys proves that this assumption is unnecessary.
% % unnecessarily conservative---when references themselves reside in TEEs, hardware isolation provides sufficient security without cryptographic overhead.
% This paper presents a fundamental observation: modern CDBs inherently rely on mapping-based protection.
Existing CDB systems using ciphertext-to-plaintext mappings suffer from critical performance bottlenecks. \sys introduces \cfmaps, which eliminate costly en/decryption from the critical path and avoid ciphertext expansion.
% \sys outperforms state of the art, and its key performance techniques have been integrated into production DBaaS systems.
The integration of \sys's key performance techniques into a production DBaaS system validates the practicality of this approach.
% Future work could explore extending crypto-free mappings to more complex scenarios: Microservices architectures with TEE-protected components could reference sensitive state through lightweight tokens rather than encrypted serialization.
% The fundamental insight---that indirection in isolation can replace encryption---provides a foundation for rethinking secure system design in the hardware-assisted confidentiality era.

\section*{Acknowledgments}

We sincerely thank our shepherd and the anonymous reviewers of OSDI 2026 for their constructive comments.
This work was supported in part by the National Natural Science Foundation of China (No. 62502510), and was sponsored by CCF-Huawei Populus Grove Fund.
Mingyu Li (\url{limingyu@ios.ac.cn}) is the corresponding author.

% \balance

\def\FEDBMAIN{}
\ifdefined\FEDBMAIN
\let\fedbappendixend\relax
\else
%%%%%%%%%%%%%%%%%%%%%%%%%%%%%%%%%%%%%%%%%%%%%%%%%%%%
% This part is included to make the appendix compilable as a standalone document.
\documentclass{article}
\usepackage[margin=1in]{geometry}
\usepackage{hyperref}
\providecommand{\sys}{ZENO}
\let\fedbappendixend\enddocument
\begin{document}
%%%%%%%%%%%%%%%%%%%%%%%%%%%%%%%%%%%%%%%%%%%%%%%%%%%%
\fi

\appendix
\section{Artifact Appendix}

\subsection*{Abstract}

The artifact for this paper contains the source code for \sys. It also includes scripts for the TPC-C and TPC-H workloads used in the paper, documentation for the two platforms described in \autoref{s:eval-setup}, and instructions for building \sys and running its smoke tests.

\subsection*{Scope}

The artifact is intended primarily to support availability and basic functionality checks. It enables users to inspect the \sys implementation, build the prototype, run the provided single-machine smoke test, and examine the scripts and documentation for the two evaluated deployment configurations: ARM S-EL2 and x86 TDX. The artifact does not include the proprietary industrial workload used in the paper.

\subsection*{Contents}

The artifact has three main parts:
\begin{enumerate}
 \item The \sys implementation in \texttt{src/}, including the components listed in \autoref{t:code-components}. The PostgreSQL integration patch is in \texttt{src/integrity\_zone/postgresql-patches/} and implements the WAL coordination path used by \sys's commit protocol. Parts of the protected-operator implementation are derived from HEDB's operator framework.
 \item Tests and benchmark scripts. \texttt{test/} contains SQL smoke tests; \texttt{benchmark/tpcc/} contains plaintext and \sys TPC-C runners based on sysbench/TPC-C Lua scripts derived from Percona code; and \texttt{benchmark/tpch/} contains TPC-H scripts derived from HEDB.
 \item Documentation in \texttt{README.md} and \texttt{doc/}. These files describe the quick-start build, single-machine simulation mode, protected-storage setup, classic two-VM deployment notes, and TD-partitioning deployment notes for x86 TDX.
\end{enumerate}

Third-party code and license notices are summarized in \texttt{THIRD\_PARTY\_LICENSES.md}.

\subsection*{Hosting}

The artifact is hosted publicly on GitHub at \url{https://github.com/ISCAS-OSLab/ZENO}. The local artifact version used to prepare this appendix corresponds to commit \texttt{213036c}.

\subsection*{Requirements}

The quick-start configuration requires no special confidential-computing hardware. In this mode, PostgreSQL and the privacy-zone proxy run as two local processes, which is sufficient for build and functional checks.

%%%%%%%%%%%%%%%%%%%%%%%%%%%%%%%%%%%%%%%%%%%%%%%%%%%%
% This part is included to make the appendix compilable as a standalone document.
\fedbappendixend
%%%%%%%%%%%%%%%%%%%%%%%%%%%%%%%%%%%%%%%%%%%%%%%%%%%%

\end{document}